\documentclass[11pt,letterpaper]{article}
\usepackage[T1]{fontenc}
\usepackage[sc]{mathpazo}
\usepackage{newunicodechar}
\newunicodechar{š}{\v{s}}
\usepackage{array}
\usepackage{microtype}
\usepackage{amsmath}
\usepackage[dvipsnames]{xcolor}
\usepackage{amssymb}
\usepackage{amsthm}
\usepackage{bm,float}
\usepackage{dsfont}
\usepackage{fullpage,caption,wrapfig}
\usepackage{graphicx}
\usepackage{comment}
\usepackage{mathtools}
\usepackage[shortlabels,inline]{enumitem}
\usepackage{bbm}
\usepackage{booktabs}
\usepackage{hyperref}
\definecolor{pku-red}{RGB}{139,0,18}
\hypersetup{
    colorlinks=true,
    linkcolor=pku-red,
    urlcolor=violet,
    citecolor=BlueViolet,
    pdffitwindow=true,
}
\AtBeginDocument{\colorlet{docnotelinkcolor}{violet}}

\makeatletter
\@ifundefined{pdfstartlink}{%
  \providecommand{\@@startlink}[1]{}%
  \providecommand{\@@endlink}{}%
}{}
\makeatother
\usepackage[capitalize, nameinlink]{cleveref}
\usepackage{nag,tikz,pgfplots}
\pgfplotsset{compat=newest}

\usetikzlibrary{arrows.meta}%
\usetikzlibrary{arrows}
\usepgfplotslibrary{colormaps,fillbetween}
\usetikzlibrary{arrows.meta, positioning, decorations.markings}

\usepackage{algorithm,algpseudocode}
\usepackage{multicol}
\usepackage[scaled]{helvet} 
\usepackage{thmtools} 
\usepackage{authblk}

\newcommand{\lineno}[1]{\noindent\hspace{-1em}{\footnotesize $#1$:\hspace{0.3em}}}

\usepackage{crossreftools}
\theoremstyle{plain}
\newtheorem{theorem}{Theorem}[section]
\newtheorem{lemma}[theorem]{Lemma}

\theoremstyle{definition}

\theoremstyle{remark}
\newtheorem{remark}[theorem]{Remark}

\usepackage{makecell}
\usepackage{subcaption}
\usepackage{tikz-cd}

\allowdisplaybreaks[4]

\newcommand{\x}{\boldsymbol{x}}
\newcommand{\ba}{\boldsymbol{a}}
\newcommand{\bb}{\boldsymbol{b}}
\newcommand{\y}{\boldsymbol{y}}
\newcommand{\s}{\boldsymbol{s}}

\newcommand{\z}{\boldsymbol{z}}
\newcommand{\e}{\boldsymbol{e}}
\newcommand{\balpha}{\boldsymbol{\alpha}}

\newcommand{\R}{\mathbb{R}}

\newcommand{\T}{\mathsf{T}}

\newcommand{\sol}{\mathsf{sol}}

\title{Discovering Expert-Level Nash Equilibrium Algorithms with Large Language Models}

\author[1]{Hanyu Li\thanks{Corresponding authors: \texttt{lhydave@pku.edu.cn}, \texttt{dongchen.li@connect.hku.hk}, \texttt{xiaotie@pku.edu.cn}}}
\author[2]{Dongchen Li{\small *}}
\author[1]{Xiaotie Deng{\small *}}
\affil[1]{CFCS, School of Computer Science, Peking University, Beijing, China}
\affil[2]{School of Computing and Data Science, The University of Hong Kong, Pokfulam, Hong Kong}

\begin{document}
\maketitle

\begin{abstract}
Designing polynomial-time algorithms for approximate Nash equilibria (ANE) with provable worst-case guarantees is a fundamental open problem in algorithmic game theory. While large language models (LLMs) can generate candidate algorithms at scale, certifying worst-case guarantees requires formal analysis over all game instances---a task for which no automated system previously existed. Here, we present LegoNE, a framework encoding expert proof strategies into a symbolic language that automatically compiles any candidate algorithm into a finite optimization problem certifying its worst-case guarantee. Integrating LegoNE with a reasoning LLM, we rediscovered an algorithm matching the best polynomial-time guarantee for two-player games, and discovered a three-player algorithm improving the best guarantee from $0.6+\delta$ to $0.5+\delta$---provably beyond the reach of the extension technique, the only previously known multi-player ANE design paradigm. These results show that encoding domain-specific proof strategies into a machine-tractable language can support LLM-driven discovery of algorithms outside known human design paradigms.

\end{abstract}

\section*{Introduction}

The Nash equilibrium (NE), introduced by Nash in 1951~\cite{nashNonCooperativeGames1951}, is a foundational solution concept in game theory. Computing an NE, however, has proven to be a persistent computational challenge: all known algorithms, such as the classic Lemke--Howson method~\cite{lemkeEquilibriumPointsBimatrix1964}, require exponential time, and the celebrated PPAD-completeness results~\cite{daskalakisComplexityComputingNash2006,chenSettlingComplexityComputing2009} confirmed that this difficulty is inherent. The study of approximate Nash equilibria (ANE) emerged in response: Lipton, Markakis, and Mehta~\cite{liptonPlayingLargeGames2003} showed that $\epsilon$-ANE can be computed in quasi-polynomial time for any constant $\epsilon$, providing the first sub-exponential algorithm; this line was further advanced by Babichenko, Barman, and Peretz~\cite{babichenkoSimpleApproximateEquilibria2014}, who reduced the dependence on the number of players from polynomial to logarithmic.

Determining the best $\epsilon$ achievable in strictly polynomial time---a canonical pursuit in the theory of approximation algorithms~\cite{williamsonDesignApproximationAlgorithms2011,vaziraniApproximationAlgorithms2003}---has been one of the fundamental open questions in algorithmic game theory~\cite{deligkasPolynomialtimeAlgorithm12023}. Rubinstein~\cite{rubinsteinSettlingComplexityComputing2016} proved that no polynomial-time algorithm can achieve $\epsilon$-ANE for a sufficiently small constant $\epsilon$ (ruling out a PTAS), though the hardness constant remains unspecified and is believed to be very small~\cite{deligkasPureCircuitTightInapproximability2024}. For two-player games with payoffs in $[0,1]$, a sequence of polynomial-time algorithms improved the worst-case guarantee from $0.75$~\cite{kontogiannisPolynomialAlgorithmsApproximating2006} to $0.5$~\cite{daskalakisNoteApproximateNash2006} to $0.3393+\delta$~\cite{tsaknakisOptimizationApproachApproximate2007}---and then progress stalled for 15 years, until Deligkas, Fasoulakis, and Markakis~\cite{deligkasPolynomialtimeAlgorithm12023} achieved $1/3+\delta$. Throughout the paper, $\delta>0$ denotes an arbitrarily small constant used in approximation guarantees; guarantees hold for any fixed $\delta$ with running time polynomial in $1/\delta$. The precise polynomial-time threshold remains unknown, with a large gap between the best algorithm and the known hardness bound.

For games with more than two players, progress has been even more limited. The only known polynomial-time approach is the \emph{extension technique}~\cite{bosseNewAlgorithmsApproximate2010,hemonApproximateNashEquilibria2008}, which recursively lifts an $r$-player algorithm to $(r{+}1)$ players with guarantee degrading from $\epsilon_r$ to $1/(2-\epsilon_r)$. Applied to the best two-player result, this yields $0.6+\delta$ for three-player games and approximately $0.71+\delta$ for four-player games, with guarantees approaching~$1$ as the number of players grows.

Large language models (LLMs) offer a route to accelerating algorithmic exploration: they can generate large numbers of candidate algorithms beyond the pace of human researchers. In recent years, combining LLMs with automated evaluators has produced advances across scientific domains, from geometry theorem proving~\cite{trinhSolvingOlympiadGeometry2024,chervonyiGoldmedalistPerformanceSolving2025} to algorithm discovery for matrix multiplication~\cite{fawziDiscoveringFasterMatrix2022} and combinatorial optimization~\cite{romera-paredesMathematicalDiscoveriesProgram2024}. Applying this paradigm to ANE algorithm design, however, requires an automated evaluator that can certify worst-case performance guarantees---that is, prove that a candidate algorithm produces a valid approximate equilibrium for every possible game instance. Such automated evaluation requires domain-specific formal tools: for Euclidean geometry, mechanized deductive systems have been available for decades~\cite{tarskiDecisionMethodElementary1951,wuMechanicalTheoremProving1994a}; for ANE algorithm analysis, no comparable automated system previously existed, and proofs have relied on ad-hoc, paper-by-paper mathematical arguments.

Here, we present LegoNE, a domain-specific framework that makes such certification automatic for ANE algorithms. LegoNE provides a symbolic language that translates expert proof strategies from two decades of ANE research into composable building blocks, and an automated analyzer that compiles any candidate algorithm written in this language into a finite optimization problem whose optimal value certifies the algorithm's worst-case guarantee. The compilation rests on two principles---\emph{instantiation} (reducing universal quantifiers to a finite set of concrete constraints) and \emph{forgetting} (abstracting payoff functions into symbolic real variables)---which together reduce an infinite-dimensional proof obligation to a fixed-size mathematical program solved by an off-the-shelf solver. Coupled with a reasoning LLM, LegoNE enables an automated discovery loop: the LLM proposes candidate algorithms by composing building blocks, the analyzer computes a provably correct approximation guarantee and returns it as quantitative feedback, and the LLM iterates toward better designs. The two components are complementary: the LLM provides large-scale exploration, and LegoNE provides rigorous, quantitative evaluation.

Using this framework, we demonstrate two main results. First, the LLM--LegoNE system rediscovered within two iterations a polynomial-time algorithm matching the best known guarantee of $1/3+\delta$ for two-player games---a result that took human experts 15 years to achieve from the previous best. Second, the system discovered a three-player algorithm achieving $0.5+\delta$, improving upon the previous best of $0.6+\delta$.

The significance of the three-player result extends beyond a numerical improvement. The extension technique---the only previously known paradigm for multi-player ANE---satisfies $\epsilon_3 = 1/(2 - \epsilon_2)$~\cite{bosseNewAlgorithmsApproximate2010,hemonApproximateNashEquilibria2008}, where $\epsilon_2$ is the two-player guarantee. Achieving $\epsilon_3 \leq 0.5$ via extension would require $\epsilon_2 = 0$---an exact Nash equilibrium, which is PPAD-hard. The discovered algorithm therefore achieves a guarantee that is provably beyond what the extension technique can deliver in polynomial time, establishing, to our knowledge, that effective multi-player ANE algorithms exist outside the extension paradigm.

These results illustrate how encoding a domain's proof strategies into a machine-tractable formal language can support LLM-driven algorithmic discovery. The high-level approach---an LLM proposes candidates while an automated evaluator provides feedback---is shared with prior AI-for-science systems~\cite{trinhSolvingOlympiadGeometry2024,romera-paredesMathematicalDiscoveriesProgram2024,fawziDiscoveringFasterMatrix2022}; our contribution is the evaluator itself, a formal system for automated worst-case analysis that did not previously exist for this domain. While LegoNE is currently specific to ANE, preliminary extensions to vertex cover approximation and polymatrix games (see the appendix) suggest that the underlying compilation technique may apply to other algorithm analysis problems where universal guarantees can be reduced to finite systems of algebraic inequalities.

\section*{Results}

\subsection*{LegoNE Framework}

LegoNE is a framework for specifying and automatically certifying algorithms for $\epsilon$-approximate Nash equilibria (ANE)~\cite{everettRecursiveGames1957}. An $\epsilon$-ANE is a near-stable strategy profile in which no player has more than $\epsilon$ regret, i.e., none can improve their payoff by more than $\epsilon$ via unilateral deviation. The smaller the worst-case $\epsilon$ guaranteed by an algorithm, the better its performance guarantee. LegoNE provides (i) a specialized symbolic language for composing ANE algorithms from high-level building blocks, and (ii) an automated analyzer that compiles any such program into a finite optimization problem whose solution certifies the algorithm's performance guarantee.

\paragraph{Language for Algorithm Design.}
The LegoNE language is a specialized Python-like language designed to specify ANE algorithms for fixed-number-of-players games. It is based on a compact set of predefined building blocks derived from established game-theoretic research over the past two decades. These blocks represent high-level strategic concepts, such as calculating an optimal counter-move (\verb`BestResponse`) or mixing existing strategies (\verb`UniformMixing`).

\begin{figure}[tbp]
    \centering
    \includegraphics[width=0.7\linewidth]{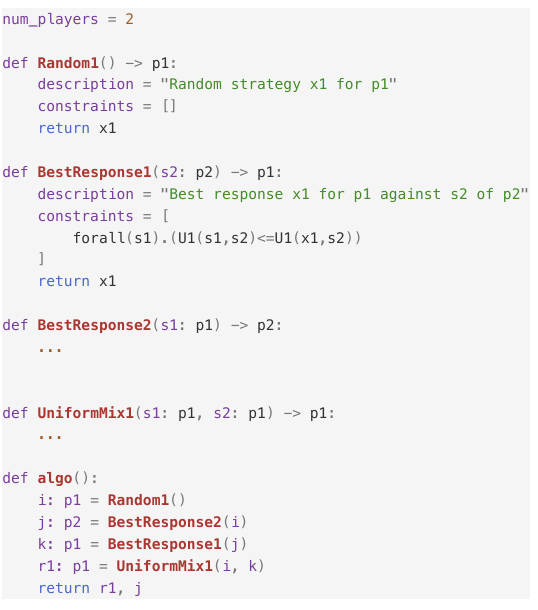}
    \caption{\textbf{The LegoNE code of the DMP algorithm.} The language uses simple, high-level building blocks encapsulating core game-theoretic concepts. Here we define building blocks \texttt{BestResponse1}, which computes the best response \texttt{x1} of player 1 (\texttt{p1}) against strategy \texttt{s2} of player 2 (\texttt{p2}), \texttt{Random1}, which randomly produces a strategy for player 1 (\texttt{p1}), and so on. Then we define the DMP algorithm, which combines these blocks and finally returns the strategy profile \texttt{r1}, \texttt{r2} for both players.}\label{figure:dmp}
\end{figure}

This modular approach simplifies algorithm design. For example, the Daskalakis-Mehta-Papadimitriou (DMP) algorithm~\cite{daskalakisNoteApproximateNash2006} can be expressed in just a few lines by composing these blocks (\Cref{figure:dmp}).

The LegoNE language is highly expressive within its domain, capable of describing a wide array of building blocks from the literature, ranging from solving linear programming to gradient descent. Moreover, it supports expressing complex algorithms through arbitrary compositions of these blocks, allowing for the creation of additional algorithms. This abstraction creates a structured design space through a programming language. Within the context, an LLM can learn to explore algorithms by combining these established concepts, rather than reasoning from scratch.

\begin{figure}[tbp]
    \centering
    \includegraphics[width=0.9\linewidth]{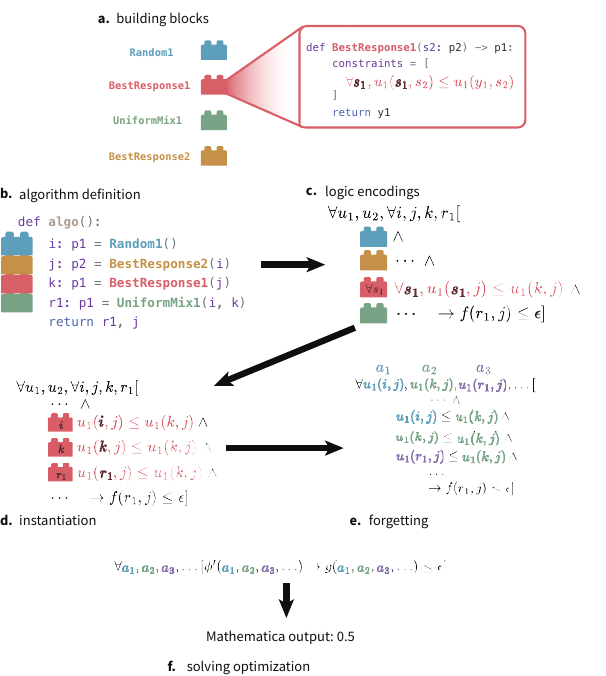}
    \caption{\textbf{The LegoNE analyzer process.} \textbf{a,} The analyzer reads all building blocks, which describe their mathematical properties. \textbf{b,} The analyzer then reads the algorithm code, which is composed of these building blocks. \textbf{c,} The analyzer translates the algorithm code into logical encodings of the algorithm's properties and the proof goal for approximation analysis. \textbf{d,} The analyzer instantiates the logical properties for the finite set of strategies constructed by the algorithm. \textbf{e,} The analyzer then ``forgets'' the underlying structure of the payoff functions and encodes the properties as a finite set of algebraic inequalities over real variables $a_1,a_2,a_3,\dots$. \textbf{f,} Finally, the analyzer formulates a constrained optimization problem to find the tightest possible $\epsilon$ subject to these inequalities and solves it using an external solver, like Mathematica.}\label{fig:analyzer-process}
\end{figure}

\paragraph{Automated Analyzer: From Code to Proof.}
A key component of LegoNE is its automated analyzer, which transforms algorithm analysis into a systematic machine-driven process. For any algorithm expressed in the LegoNE language, the analyzer computes its best possible approximation guarantee, $\epsilon$, and simultaneously generates a computer proof of this guarantee. This is achieved through a two-step abstraction that reduces the original infinite-dimensional problem which states a guarantee for all possible games to a finite, solvable math program. An overview of this process is illustrated in \Cref{fig:analyzer-process}.

The process begins by translating the procedural code of an algorithm into a set of declarative logical properties based on Floyd-Hoare semantics~\cite{floydAssigningMeaningsPrograms1993,hoareAxiomaticBasisComputer1969}. During the process, each line of the code is encoded as a mathematical statement. For instance, the instruction 
\begin{center}
\texttt{k = BestResponse1(j)}
\end{center}
becomes a universal assertion: ``Given that player 2 plays strategy \verb`j`, for all possible strategies player 1 could choose, none yields a higher payoff than strategy \verb`k`.'' More formally, this can be expressed as a logical statement as follows:
\begin{equation}\forall s_1, u_1(s_1,j) \le u_1(k,j),\end{equation}
where $u_1(s_1,j)$ is the payoff for player 1 when playing strategy $s_1$ against player 2's strategy $j$. This statement asserts that for all possible strategies $s_1$ available to player 1, none yields a higher payoff than strategy \verb`k` when player 2 plays strategy \verb`j`. Aggregating these properties for all lines of code creates a complete series of logical properties of the algorithm.

Because this specification ranges over infinite-dimensional spaces (the set of all strategies and payoff functions), direct computation is intractable; the core difficulty is universal quantification (e.g., $\forall s_1$ over all strategies). LegoNE addresses this via a two-step process: instantiation and forgetting.

\begin{enumerate}
    \item \textbf{Instantiation: From Infinite to Finite.} The analyzer instantiates the logical properties only for the finite set of strategies explicitly constructed within the algorithm. This is based on the insight from human proofs that an algorithm's properties need only be verified for the strategies it generates. For example, to analyze the line 
    \begin{center}
    \texttt{k = BestResponse1(j)},
    \end{center}
    which corresponds to the property $\forall s_1, u_1(s_1, j) \le u_1(k, j)$, the analyzer instantiates the universal quantifier $\forall s_1$ with exactly the strategies for player 1 used in the DMP algorithm (\Cref{figure:dmp}): \verb`i`, \verb`k`, and \verb`r1`. This substitution yields a finite set of inequalities involving only the strategies that the algorithm actually constructs:
    \begin{itemize}
        \item $u_1(i, j) \le u_1(k, j)$
        \item $u_1(k, j) \le u_1(k, j)$ (which is trivially true)
        \item $u_1(r_1, j) \le u_1(k, j)$
    \end{itemize}
    This step reduces the infinite set of quantifiers to a finite number of instantiated constraints.

    \item \textbf{Forgetting: From Functions to Variables.} Although the number of constraints is now finite, terms like $u_1(i, j)$ still depend on unknown payoff functions from arbitrary games. To address this second source of infinitude, the analyzer treats each payoff value, such as $u_1(i, j)$, as a single symbolic real variable, denoted $v_{1,ij}$. This forgets the underlying structure of the function $u_1$. The inequality $u_1(i, j) \le u_1(k, j)$ is thus transformed into a simple algebraic relationship between two variables: $v_{1,i,j} \le v_{1,k,j}$.
\end{enumerate}

Through this two-step process, the initial assertion over infinite-dimensional functions is converted into a finite system of algebraic inequalities over a set of real variables. This system is then formulated as a constrained optimization problem: find the minimum possible $\epsilon$ --- an expression also formulated using these new variables --- subject to the derived system of inequalities. The optimal value, found by an external solver, is the tightest possible worst-case approximation guarantee, $\epsilon$. The solution to this optimization problem constitutes a constructive proof of the algorithm's performance guarantee. Within the LegoNE framework, computing the guarantee is equivalent to proving it.

\paragraph{Empirical Validation.}
To verify the correctness of the LegoNE analyzer, we implemented all known polynomial-time ANE algorithms for fixed-number-of-players games from the literature in LegoNE language. This includes a series of works published over more than two decades. The literature on ANE algorithms has focused almost exclusively on two-player games. For games with more than two players, the only established design paradigm is to extend two-player algorithms. We also implemented the three-player algorithm derived from this paradigm.

The result is shown in \Cref{tab:benchmark}. For each algorithm, the LegoNE analyzer managed to compute the approximation guarantee matching the results from the original papers up to a $10^{-5}$ precision. The original proofs for these algorithms required from a few to over a dozen pages of mathematical arguments. In LegoNE, each algorithm was expressed within 60 lines of code. The automated computation of the approximation guarantee for each was completed within 80 seconds for all algorithms, a task that had previously required years of cumulative human research. These results confirmed the correctness of the framework and showed its potential to accelerate theoretical research by efficiently identifying correct approximation guarantees. See \Cref{tab:benchmark-detail} for the detailed experimental results.

\begin{table}[tbp]
    \centering
    \begin{tabular}{c|cc}
        \toprule
        \textbf{Author initials, year} & \makecell{\bfseries Guarantee proved in\\\bfseries original paper} & \makecell{\bfseries Guarantee proved\\\bfseries by LegoNE} \\
        \midrule
        KPS~\cite{kontogiannisPolynomialAlgorithmsApproximating2006}, 2006 & $0.75$ & $0.75000$ \\
        DMP~\cite{daskalakisNoteApproximateNash2006}, 2006 & $0.5$ & $0.50000$ \\
        DMP~\cite{daskalakisProgressApproximateNash2007}, 2006 & $0.38197+\delta$\textsuperscript{\textdagger} & $0.38197+\delta$ \\
        BBM-1~\cite{bosseNewAlgorithmsApproximate2010}, 2007 & $0.38197$ & $0.38197$ \\
        CDFFJS~\cite{czumajDistributedMethodsComputing2016}, 2016 & $0.38197$ & $0.38197$ \\
        BBM-2~\cite{bosseNewAlgorithmsApproximate2010}, 2007 & $0.36392$ & $0.36392$ \\
        TS~\cite{tsaknakisOptimizationApproachApproximate2007}, 2007 & $0.33933+\delta$ & $0.33933+\delta$ \\
        DFM~\cite{deligkasPolynomialtimeAlgorithm12023}, 2022 & $1/3+\delta$ & $0.33333+\delta$ \\
        \makecell{DFM+extension (3-player)\\\cite{deligkasPolynomialtimeAlgorithm12023,bosseNewAlgorithmsApproximate2010,hemonApproximateNashEquilibria2008}, 2022} & $0.6+\delta$ & $0.60000+\delta$ \\
        \bottomrule
    \end{tabular}
    \caption{\textbf{Benchmark of LegoNE on existing algorithms.} The table summarizes the approximation guarantees proved by LegoNE for each algorithm, compared to the guarantees proven in the original papers. The results match exactly, demonstrating the framework's correctness and effectiveness in automating algorithm analysis. \textsuperscript{\textdagger}Throughout the paper, $\delta>0$ denotes an arbitrarily small constant used in approximation guarantees; guarantees hold for any fixed $\delta$ with running time polynomial in $1/\delta$.}\label{tab:benchmark}
\end{table}

\begin{table}[tbp]
    \centering
    \begin{tabular}{c|ccc}
        \toprule
        \textbf{Author initials, year} & \makecell{\bfseries Guarantee proved\\\bfseries by LegoNE} & \makecell{\bfseries Code lines} & \makecell{\bfseries Running time (s)} \\
        \midrule
        KPS~\cite{kontogiannisPolynomialAlgorithmsApproximating2006}, 2006 & $0.75000$ & 27 & 22.55 \\
        DMP~\cite{daskalakisNoteApproximateNash2006}, 2006 & $0.50000$ & 42 & 4.31\\
        DMP~\cite{daskalakisProgressApproximateNash2007}, 2006 & $0.38197+\delta$ & 38 & 31.13 \\
        BBM-1~\cite{bosseNewAlgorithmsApproximate2010}, 2007 & $0.38197$ & 46 & 11.81 \\
        CDFFJS~\cite{czumajDistributedMethodsComputing2016}, 2016 & $0.38197$ & 48 & 65.87 \\
        BBM-2~\cite{bosseNewAlgorithmsApproximate2010}, 2007 & $0.36392$ & --- & 1.38 \\
        TS~\cite{tsaknakisOptimizationApproachApproximate2007}, 2007 & $0.33933+\delta$ & 31 & 13.90 \\
        DFM~\cite{deligkasPolynomialtimeAlgorithm12023}, 2022 & $0.33333+\delta$ & 52 & 79.63 \\
        \makecell{DFM+extension (3-player)\\\cite{deligkasPolynomialtimeAlgorithm12023,bosseNewAlgorithmsApproximate2010,hemonApproximateNashEquilibria2008}, 2022} & $0.60000+\delta$ & 50 & 8.34 \\
        \bottomrule
    \end{tabular}
    \caption{\textbf{Detailed benchmark of LegoNE on existing algorithms.} This table shows the lines of code and running time for each algorithm. The code lines include defining all used building blocks, inherent constraints, and the algorithm definition. The guarantees computed by LegoNE are also listed for reference.}\label{tab:benchmark-detail}
\end{table}

\paragraph{Extending the Framework.}
The framework's core principles, instantiation and forgetting, extend beyond the analysis of fixed-number-of-players games. We have applied the framework to analyze algorithms for two broader classes of problems, demonstrating its generality. See the appendix for details.

First, we analyzed algorithms for approximate Nash equilibrium in polymatrix games~\cite{deligkasComputingApproximateNash2017}. Unlike standard games with a fixed number of players, polymatrix games model interactions in large networks where each player's payoff is determined only by their neighbors in a given graph. The challenge is to reason about an arbitrary number of players. We applied our framework to the only known polynomial-time ANE algorithm for polymatrix games from the literature. Our framework handles the analysis by using the forgetting principle to abstract away player-specific indices, allowing for a unified analysis that automatically proves the algorithm's approximation guarantee.

Second, we extended the framework to analyze a broad class of approximation algorithms based on linear programming (LP) relaxation and rounding~\cite{vaziraniApproximationAlgorithms2003}. As a classic example of this method, we analyzed an algorithm for the vertex cover problem, a fundamental challenge in computer science. The analysis involves translating the algorithm --- which solves a relaxed version of the problem and rounds the solution --- into our framework. The analyzer then automatically proves the algorithm's approximation ratio of $2$, matching the known result from the literature. These applications show that the framework can potentially serve as a general tool for automated algorithm analysis across a broader range of computational problems.

\subsection*{LLM-Powered Algorithmic Discovery}

\begin{figure}[tbp]
    \centering
    \includegraphics[width=0.8\linewidth]{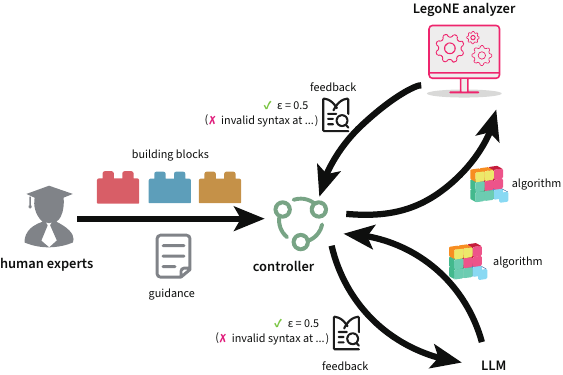}
    \caption{\textbf{The human-machine collaborative loop for algorithm discovery.} Human experts provide high-level building blocks and design guidance. The LLM then proposes a candidate algorithm by combining these blocks. The LegoNE analyzer automatically computes a proven approximation guarantee and returns it as feedback; if the algorithm has a syntax error, the analyzer will output the specific error message to the LLM. The loop iterates, allowing the LLM to refine its designs based on rigorous theoretical feedback.}\label{fig:loop}
\end{figure}

We further integrate LegoNE with an LLM to automate algorithmic discovery (\Cref{fig:loop}).

In this paradigm, human experts and LLMs have different roles based on their complementary strengths. The human expert provides the fundamental components for algorithm design, translating domain-specific knowledge and high-level proof strategies into the building blocks of the LegoNE language. This task requires a level of conceptual understanding not yet achievable by LLMs. The LLM, in turn, navigates the vast combinatorial space of how these blocks can be assembled, exploring potential algorithms at a scale and speed difficult for human researchers alone.

The learning process in this framework is useful for two reasons. First, the building blocks provide a high-level specification of the design space, making it tractable for the LLM. Second, the LegoNE analyzer provides rapid, rigorous feedback by computing a provably correct approximation guarantee for any proposed algorithm. This iterative process of proposing, verifying, and refining allows for efficient exploration of the algorithm design space. An interaction controller manages this process, using carefully designed prompts from human experts' design experience and insights to guide the LLM toward promising areas of the design space while still encouraging the generation of diverse solutions. Importantly, this feedback is not evaluated on particular game instances: the analyzer returns an approximation guarantee that holds for all game instances.

\paragraph{Reproducing Best Polynomial-Time Worst-Case Guarantees.}
To validate the framework, we first tasked it with rediscovering a known result in ANE research. In an experiment configured for two-player games, a reasoning LLM~\cite{guoDeepSeekR1IncentivizesReasoning2025} was provided only with the building blocks available in the literature prior to 2007, when the best polynomial-time worst-case guarantee at the time~\cite{tsaknakisOptimizationApproachApproximate2007} was obtained. After 2 rounds of interaction with the LegoNE analyzer, the LLM constructed an algorithm. Although structurally different from the 2022 refinement achieving the current best polynomial-time worst-case guarantee~\cite{deligkasPolynomialtimeAlgorithm12023}, LegoNE proved it achieved the same approximation guarantee. This result, which required 15 years of cumulative research by human experts, indicates the framework's potential to accelerate the process of generating and validating theoretical ideas.

\begin{figure}[tbp]
    \centering
    \includegraphics[width=0.9\linewidth]{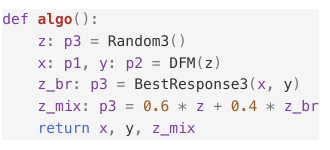}
    \caption{\textbf{An extension-technique three-player ANE algorithm (baseline).} This construction treats the best two-player polynomial-time NE algorithm as a black-box component and extends it to three-player games, yielding the previously best $0.6+\delta$ polynomial-time worst-case guarantee.}\label{figure:extension3p}
\end{figure}

\paragraph{Discovering Multi-Player Algorithms.}
We applied the framework to three-player games, a setting where human-designed methods have seen limited progress. Prior work typically relies on an extension technique that adapts two-player algorithms~\cite{bosseNewAlgorithmsApproximate2010,hemonApproximateNashEquilibria2008}; this approach is restrictive and often yields weaker guarantees. \Cref{figure:extension3p} illustrates the three-player polynomial-time baseline that, prior to our work, achieved the best polynomial-time worst-case guarantee, constructed using the extension technique over the DFM algorithm~\cite{deligkasPolynomialtimeAlgorithm12023}.

\begin{figure}[tbp]
    \centering
    \includegraphics[width=\linewidth]{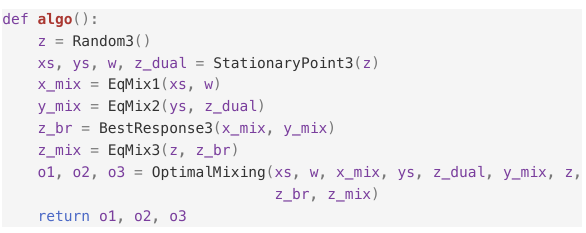}
    \caption{\textbf{A three-player ANE algorithm discovered by the LLM.} This algorithm combines building blocks such as \texttt{StationaryPoint3} and \texttt{OptimalMixing} in a distinctive structure. The LegoNE framework proves an approximation guarantee of $0.5+\delta$, improving upon the previously best $0.6+\delta$ polynomial-time worst-case guarantee.}\label{figure:new3p}
\end{figure}

Using the LLM-LegoNE system, we discovered in 11 rounds a fundamentally different algorithm (\Cref{figure:new3p}) whose LegoNE analysis proves an approximation guarantee of $0.5+\delta$, improving on the previously best $0.6+\delta$ polynomial-time worst-case guarantee.

The central building block in \Cref{figure:new3p} is \texttt{StationaryPoint3}. Intuitively, it fixes player 3's mixed strategy $z$ and performs a gradient descent procedure using linear programming to find a stationary point $(x_s,y_s)$ that locally minimizes the maximum regrets of players 1 and 2, associated with dual strategies $(w,z_{\text{dual}})$.

Beyond this numerical improvement, the discovery represents a departure from the only previously known design paradigm for multi-player ANE. The extension technique satisfies $\epsilon_3 = 1/(2-\epsilon_2)$~\cite{bosseNewAlgorithmsApproximate2010,hemonApproximateNashEquilibria2008}; achieving $\epsilon_3 \leq 0.5$ via extension would require $\epsilon_2 = 0$---an exact Nash equilibrium, which is PPAD-hard. The discovered algorithm therefore achieves a guarantee that no extension-based construction can deliver in polynomial time. Structurally, while extension-technique constructions use a best-response-then-mix approach as the final step, the discovered algorithm uses a different composition of building blocks, combining \texttt{BestResponse}, \texttt{EqMix}, and \texttt{OptimalMixing}.

Across 12 rounds, the LLM produced 7 distinct algorithms; only the first relied on the extension technique, while the remaining 6 explored alternative non-extension structures that nonetheless achieved nontrivial guarantees. The expert knowledge provided to the LLM consists of the building blocks themselves (standard game-theoretic operations drawn from the ANE literature, such as \texttt{BestResponse}, \texttt{StationaryPoint}, and \texttt{OptimalMixing}) and prompt-level constraints that define the search space---but not the algorithmic structure, composition order, or mixing coefficients. In these runs, the extension technique was available as a possible component, yet the LLM generated non-extension approaches. The structure of the final $0.5+\delta$ algorithm was not anticipated by the human experts who designed the building blocks. A detailed inventory of the expert inputs is provided in the Methods (Section~B.1).

\section*{Discussion}

This work presents LegoNE, a domain-specific formal system that automates the analysis of approximate Nash equilibrium algorithms, and demonstrates its integration with an LLM to form an automated discovery loop. The LegoNE analyzer---built on the instantiation and forgetting principles---compiles any candidate algorithm expressed in the LegoNE language into a finite optimization problem whose solution certifies the algorithm's worst-case guarantee. Coupled with a reasoning LLM, this loop reproduced the best polynomial-time worst-case guarantee for two-player games within two iterations and discovered a three-player algorithm that improves upon the previous best guarantee from $0.6+\delta$ to $0.5+\delta$.

The three-player result has implications beyond the specific approximation constant. The extension technique~\cite{bosseNewAlgorithmsApproximate2010,hemonApproximateNashEquilibria2008}---previously the only known paradigm for designing multi-player ANE algorithms---satisfies $\epsilon_3 = 1/(2-\epsilon_2)$; achieving $\epsilon_3 \leq 0.5$ via extension would require an exact Nash equilibrium algorithm, a PPAD-hard task. The discovered algorithm therefore establishes, to our knowledge, that effective polynomial-time multi-player ANE algorithms exist outside the extension paradigm. This opens a design space for multi-player algorithms and contributes to narrowing the gap between the best known polynomial-time guarantees and the hardness lower bounds---a gap that remains large and largely uncharacterized for both two-player and multi-player settings.

Our progress should be read as both a positive result and a boundary marker. LegoNE currently depends on human-curated building blocks that encode proof strategies from the existing literature; it cannot discover fundamentally different proof techniques. The framework's scope is limited to algorithm analysis problems where the instantiation and forgetting principles apply---settings where universal guarantees can be reduced to finite systems of algebraic inequalities. Our preliminary extensions to vertex cover approximation and polymatrix games (see the appendix) suggest this class is nontrivial, but its exact boundaries remain to be characterized. Looking ahead, we hope to develop more expressive languages that allow algorithms to be analyzed from first principles, reducing the dependence on hand-curated building blocks and broadening the range of problems amenable to automated analysis.

\clearpage

\section*{Methods}
\section*{The LegoNE Framework in Detail}\label{sec:legone-detail}

\subsection*{Background: Approximate Nash Equilibrium (ANE)}

In a normal-form game, multiple players select actions from their respective action spaces. Players can choose mixed strategies, which are probability distributions over their available actions, and they subsequently receive expected payoffs. Suppose there are $r$ players, with $r$ fixed. A strategy profile, denoted by $\x = (\x^1, \dots, \x^r)$, is a collection of all players' strategies. Each player $i$ has a payoff function, $u_i(\x)$, which represents their expected utility under a given strategy profile $\x$. Following the standard convention for approximation guarantees, these payoffs are normalized to the range $[0, 1]$~\cite{kontogiannisPolynomialAlgorithmsApproximating2006}. A central concept in game theory is the Nash equilibrium (NE), a strategy profile where no single player can gain a better payoff by unilaterally changing their strategy. The existence of at least one NE is guaranteed in any finite game~\cite{nashNonCooperativeGames1951,nashEquilibriumPointsNperson1950}.

While the concept of NE is fundamental, the focus of modern algorithmic game theory often shifts to the notion of an \emph{approximate Nash equilibrium} (ANE)~\cite{everettRecursiveGames1957}, which relaxes the strict optimality condition of an NE. The regret for player $i$ under a strategy profile $\x$, denoted $f_i(\x)$, measures the maximum payoff increase they could achieve by deviating to a different strategy. Formally, this regret is defined as:
\begin{equation}
f_i(\x) := \max_{\tilde{\x}^i\in\Delta_{n_i}}u_i(\tilde{\x}^i, \x^{-i}) - u_i(\x) .
\end{equation}
Here, $\x^{-i}$ represents the strategies of all players except player $i$. For any non-negative value $\epsilon$, a strategy profile $\x$ is termed an $\epsilon$-approximate Nash equilibrium ($\epsilon$-NE) if the maximum regret among all players does not exceed $\epsilon$; that is, if $f(\x) := \max_{i\in[r]}f_i(\x) \leq \epsilon$. A perfect Nash equilibrium corresponds to the case where $\epsilon=0$. 

When we say an algorithm has an approximation guarantee of $b$, we mean that it produces a strategy profile $\x$ such that $f(\x) \leq b$ for any game instance. Consequently, the objective for designing an ANE algorithm is to create a polynomial-time process whose resulting strategy profile $\x$ consistently ensures a small $f(\x)$ value for any game instance. We focus on algorithms for a fixed number $r$ of players.

\subsection*{The LegoNE Language}

The LegoNE framework introduces a domain-specific programming language that shifts the paradigm from describing an algorithm's execution steps (its \emph{operational semantics}~\cite{maillouxReportAlgorithmicLanguage1969}) to specifying its mathematical properties (its \emph{axiomatic semantics}~\cite{floydAssigningMeaningsPrograms1993,hoareAxiomaticBasisComputer1969}). This methodology is inspired by Floyd-Hoare logic~\cite{floydAssigningMeaningsPrograms1993,hoareAxiomaticBasisComputer1969}, and facilitates a human-machine collaboration where human designers encode their insights about algorithmic components into logical formulas, which machines then use as axioms for further analysis.

An algorithm is constructed by first defining fundamental \emph{building blocks} and then composing them in sequence. Consider the Daskalakis-Mehta-Papadimitriou (DMP) algorithm~\cite{daskalakisNoteApproximateNash2006} for $2$-player games. Its building blocks include operations like \verb|BestResponse|, which finds an optimal strategy against an opponent (e.g., $\boldsymbol{j} = \mathsf{BestResponse}(\boldsymbol{i})$), and \verb|UniformMixing|, which creates an equal-probability mixture of two strategies (e.g., $\boldsymbol{r}_1 = \mathsf{UniformMixing}(\boldsymbol{i}, \boldsymbol{k})$). Although these examples are simple, the LegoNE language is highly expressive within its domain, capable of describing a wide array of building blocks from the literature, ranging from solving linear programming to gradient descent.

The core of LegoNE is that each such building block is encoded by a logic formula that captures its guaranteed properties. For instance, the \verb|BestResponse| operation $\boldsymbol{j} = \mathsf{BestResponse}(\boldsymbol{i})$ for player 2 is not described by how to compute it, but by the logical assertion that it produces a payoff at least as high as any other possible strategy $\boldsymbol{y}$:
\begin{equation}\forall \boldsymbol{y}(u_2(\boldsymbol{i}, \boldsymbol{y}) \leq u_2(\boldsymbol{i}, \boldsymbol{j})).\end{equation}
Similarly, the \verb|UniformMixing| block is encoded with formulas specifying its linear behavior on payoffs, such as $\forall \boldsymbol{y}(u_1(\boldsymbol{r}_1, \boldsymbol{y}) = u_1(\boldsymbol{i}, \boldsymbol{y})/2 + u_1(\boldsymbol{k}, \boldsymbol{y})/2)$.

The logical encoding for an entire algorithm, $\phi[\Gamma]$, is the conjunction of the formulas for each of its steps, along with a set of predefined \emph{inherent formulas}, $\phi_0$. These inherent formulas state fundamental truths, such as the definition of regret and the fact that all payoffs are bounded within $[0, 1]$. The objective of proving that algorithm $\Gamma$ has an approximation guarantee of $\epsilon$ is then stated as a single, comprehensive logical implication:
\begin{equation}(\forall u_1, \dots, u_r) (\forall s_{11}, \dots) (\phi[\Gamma] \to f(\boldsymbol{s}^1, \dots, \boldsymbol{s}^r) \leq \epsilon).\end{equation}
This formula asserts that for any valid game, if the properties guaranteed by the algorithm's steps hold, then its output's $f$ value is necessarily bounded by $\epsilon$; thus, the approximation guarantee of this algorithm is $\epsilon$.

\subsection*{The LegoNE Analyzer}

The central challenge in formally analyzing an ANE algorithm lies in handling objects of unbounded dimensionality. An algorithm's approximation guarantee must hold true for any game. This implies it must be valid for payoff matrices of any arbitrary size and for an infinite continuum of possible mixed strategies within the corresponding simplexes. This universal-quantification nature of the proof goal makes direct computational verification intractable.

To overcome this fundamental obstacle, the LegoNE analyzer introduces a two-step procedure that systematically translates the abstract, infinite-dimensional proof task into a concrete, fixed-size constrained optimization problem. This procedure is inspired by how human experts construct proofs: by separating the properties of the algorithm from the general properties of real arithmetic. The core of our automated analysis rests on two principles: \emph{instantiation} and \emph{forgetting}.

\paragraph{The Instantiation Principle: From Infinite Quantifiers to Finite Constraints.}
LegoNE's first key tactic, instantiation, translates the universally quantified properties of an algorithm into a finite set of concrete algebraic inequalities. A human proof does not reason about all infinitely many strategies; instead, it cleverly selects a few crucial instances to build its argument. For example, a building block like $\x^k = \mathsf{BestResponse}(\x^j)$ is defined by the property $\forall \x (u_1(\x, \x^j) \leq u_1(\x^k, \x^j))$. The analyzer automates the human-like instantiation process by systematically substituting the universally quantified variable (here, $\forall \x$) with all other specific strategy variables that appear in the algorithm's code.

Furthermore, it instantiates universally quantified variables using a strategy that maximizes the payoff. For example, from $\forall \x (u_1(\x, \x^j) \leq u_1(\x^k, \x^j))$ we can derive $\max_{\x} u_1(\x, \x^j) \leq u_1(\x^k, \x^j)$. This allows it to derive a crucial identity, for instance, $\max_{\x} u_1(\x, \x^j) = u_1(\x^k, \x^j)$, by combining the derived inequality $\max_{\x} u_1(\x, \x^j) \leq u_1(\x^k, \x^j)$  with the inherent property from the definition of the maximum operator, $\max_{\x} u_1(\x, \x^j) \geq u_1(\x^k, \x^j)$. By applying this instantiation procedure to every logical formula describing the algorithm, LegoNE transforms the problem from one of infinite logical deduction to one of satisfying a finite system of inequalities involving a limited number of strategy-dependent terms.

\paragraph{The Forgetting Principle: From Function Spaces to Real Variables.}
After instantiation, the problem is reduced to a system of inequalities involving terms like $u_1(\x^i, \x^j)$, $u_1(\x^k, \x^j)$, and $\max_{\x} u_1(\x, \x^j)$. These terms are still, in principle, complex, high-dimensional payoff functions. The second key tactic, forgetting, is based on the insight that a formal proof only uses the arithmetic relationships between these terms, not their underlying functional structure.

Therefore, the analyzer forgets their origin. It treats each unique term, such as $u_1(\x^i, \x^j)$, as a single, abstract real-valued variable, for example, $a_1$. The term $u_1(\x^k, \x^j)$ becomes $a_2$, $\max_{\x} u_1(\x, \x^j)$ becomes $a_3$, and so on for all terms present in the instantiated formula. The original system of inequalities is thus rewritten purely in terms of these new real variables. This crucial step maps the problem from an intractable function space into a standard, finite-dimensional real vector space, where only the core arithmetic properties that are essential for the proof are preserved.

\paragraph{The End Product: A Constrained Optimization Problem.}
The culmination of the instantiation and forgetting principles is the automatic compilation of any algorithm written in LegoNE into the following form:
\begin{equation}\forall a_1, a_2, \dots (\phi'(a_1, a_2, \dots) \to g(a_1, a_2, \dots) \leq \epsilon),\end{equation}
where $\phi'(a_1, a_2, \dots)$ is a set of inequalities derived from the algorithm's properties, and function $g(a_1, a_2, \dots)$ represents the original function $f(\x)$, now expressed in terms of the abstract variables $a_1, a_2, \dots$. The goal is to find the minimum value of $\epsilon$ such that the implication holds for all possible values of the abstract variables. This is equivalent to finding the maximum value of $g(a_1, a_2, \dots)$ subject to the constraints $\phi'(a_1, a_2, \dots)$. The formulation is as follows:
\begin{equation}\begin{aligned}
    \operatorname*{maximize}_{a_1, a_2, \dots} &\quad g(a_1, a_2, \dots)\\
    \text{subject to} &\quad\phi'(a_1, a_2, \dots).
\end{aligned}\end{equation}
The optimal value of this fixed-size optimization problem is precisely the best approximation guarantee for the given algorithm that can be derived via this proof tactic. Crucially, the number of variables and constraints of this optimization problem is fixed once the tactic is fixed and does not scale with the size of the game instance. This final problem is a fixed-size program that can be solved by off-the-shelf numerical solvers, such as Gurobi~\cite{GurobiOptimization} or Mathematica~\cite{WolframMathematicaModern}.

\subsection*{Benchmarking LegoNE on Existing Human-Designed Algorithms}\label{subsec:benchmark-existing}

To validate the correctness and analytical power of the LegoNE framework, we benchmarked it against the existing human-designed ANE algorithms from the literature. The framework's compiler was implemented in C++ using the lexer Flex~\cite{paxsonFlexScannerGenerator} and parser Bison~\cite{freesoftwarefoundationBisonGNUProject}, with Mathematica serving as the external optimization solver. The version of Wolfram Engine inside Mathematica is 14.2. We set the Mathematica parameters as follows: \verb|AccuracyGoal| to 10, \verb|WorkingPrecision| to 20, and \verb|MaxIterations| to 2000. We use the default method (auto) as the optimization method. To ensure a reproducible result, we rerun the same optimization on a MacBook and a Windows PC. For testing the running time, we used a MacBook Pro 14 (M4 chip, 10-core CPU/10-core GPU, 16GB RAM). Each optimization was executed 5 times under identical conditions.

Most existing polynomial-time ANE algorithms have been designed for two-player games. We encoded all major algorithms from the literature in the LegoNE language, including complex ones like the Tsaknakis-Spirakis algorithm. This process demonstrated LegoNE's ability to express sophisticated algorithmic components, such as branching operations, mixing strategies with payoff-dependent coefficients, and the intricate \texttt{StationaryPoint} building block. 

For computational efficiency and due to current limitations of the LegoNE compiler, we set special parameters for some algorithms. The LegoNE compiler does not currently support branching operations; however, due to symmetry of the algorithm, it is sufficient to encode one branch. For BBM-1~\cite{bosseNewAlgorithmsApproximate2010} with $0.36$ approximation guarantee, the LegoNE compiler does not currently support square root operations, so we manually compute the compiler output. For KPS~\cite{kontogiannisPolynomialAlgorithmsApproximating2006} with $0.75$ approximation guarantee and DMP~\cite{daskalakisProgressApproximateNash2007} with $0.38$ approximation guarantee, we manually provide the diagonal logic encoding (see the appendix remark on optimal mixing for two-player games) for the optimal mixing operation to follow the original proof.

As summarized in \Cref{tab:benchmark,tab:benchmark-detail}, the approximation guarantees computed and proven by LegoNE for each algorithm are identical (up to $10^{-5}$ precision) to the guarantees proven in the original papers.

To further demonstrate the framework's expressive power, we modeled the only established paradigm for designing three-player ANE algorithms in the literature: the extension technique~\cite{bosseNewAlgorithmsApproximate2010,hemonApproximateNashEquilibria2008}. This method takes an existing algorithm $\Gamma$ that computes an $\alpha$-NE for two-player games and extends it to create an algorithm for three-player games. Prior to our work, the best human-designed polynomial-time worst-case guarantee for three-player games was obtained by extending the $(1/3+\delta)$-NE two-player algorithm~\cite{deligkasPolynomialtimeAlgorithm12023}, which yields a $(0.6+\delta)$-NE. We modeled this in LegoNE by treating the two-player algorithm $\Gamma$ as a single, black-box building block whose encoding is its proven guarantee (e.g., $f(\x^1_o,\x^2_o)\leq 0.6+\delta$). When analyzing the complete three-player algorithm, LegoNE correctly computed the approximation guarantee to be $(0.6+\delta)$, matching the result proven in the literature.

\section*{Experimental Details for Automated Algorithmic Discovery}

\subsection*{Human-Machine Interaction Loop and Engineering Considerations}

Our automated discovery process is centered on an iterative interaction loop between a Large Language Model (LLM) and the LegoNE framework. This approach is founded on a human-machine collaboration paradigm where humans establish the high-level theoretical framework, and the machine explores the vast algorithmic design space within it. The process separates the definition of foundational building blocks from the task of combining them into a complete algorithm. Defining additional building blocks requires deep domain expertise and complex mathematical derivations, a role reserved for human experts. In contrast, the task of combining these predefined blocks is a more structured, mechanical process that is well-suited for automation by an LLM. This automation is feasible due to two key factors: the design space is significantly condensed by using complex, pre-defined blocks that encapsulate human insights, and the LegoNE compiler provides immediate feedback by automatically computing the approximation guarantee for any valid combination.

To make this exploration process efficient and robust, we implemented several engineering choices:

\paragraph{Auto-return with Optimal Mixing}
In the literature of ANE algorithms, a particularly complex step is determining the appropriate coefficients for mixing constructed strategies to achieve a small approximation guarantee. This task is challenging for LLMs. To simplify the task, we modified the LegoNE compiler to automatically apply optimal mixing operations as the final step of any given algorithm fragment. This enhancement allows the LLM to focus on constructing effective strategies, while the compiler handles the optimal mixing.

\paragraph{Prompt Engineering}
We designed a multi-stage prompting strategy to guide the LLM.
\begin{itemize}
    \item \textbf{Initial Prompt:} The first prompt provides a clear and constrained starting point for the LLM. It includes: 
    \begin{enumerate*}[label=(\arabic*)]
        \item a clear description of the task, with the primary goal of minimizing the approximation guarantee $\epsilon$;
        \item strict design constraints, such as the mandatory use of static single assignments (SSA), type annotations, a no-return statement requirement, and the compulsory inclusion of at least one $\mathsf{StationaryPoint}$ building block due to its proven effectiveness from human experts' experience; and
        \item concrete examples of both valid and invalid code snippets to help the LLM understand the expected format and constraints.
    \end{enumerate*}

    \item \textbf{Iterative Feedback Prompts:} Subsequent prompts are dynamically adjusted based on the LLM's previous output.
        If the previous algorithm was invalid, the new prompt includes detailed compiler error messages, highlighting specific issues like type mismatches, incorrect usage of building blocks, or violations of SSA rules. The LLM is then instructed to carefully review and correct these errors.
        If the algorithm was valid, the prompt focuses on optimization. It provides the current best approximation guarantee $\epsilon$ and encourages the LLM to achieve a smaller value. Specifically, the prompt guides the LLM to: leverage more complex building blocks; break symmetry by applying different blocks to each player; explore alternative combinations to avoid local optima; and adhere to Occam's Razor by preferring simpler strategies that achieve comparable or better results.
\end{itemize}

\paragraph{Expert Knowledge Inventory}
To enable assessment of whether the expert-provided knowledge may have biased the LLM toward the discovered solutions, we provide a detailed accounting of all expert inputs.

The building blocks provided to the LLM encode standard game-theoretic operations drawn from the ANE literature:
\begin{itemize}
    \item \texttt{BestResponse}: computes an optimal response to an opponent's strategy (basic game-theoretic concept);
    \item \texttt{Random}: generates an arbitrary strategy (trivial);
    \item \texttt{StationaryPoint}: computes an approximate stationary point of the maximum-regret function and its dual, returning four strategies~\cite{tsaknakisOptimizationApproachApproximate2007};
    \item \texttt{ZeroSumNE}: computes an exact Nash equilibrium of a zero-sum game via linear programming (textbook);
    \item \texttt{EqMix}: equal-probability mixture of two strategies (trivial);
    \item \texttt{MaxPayoff}: maximizes a given payoff function (trivial).
\end{itemize}
These building blocks define what operations are available, but do not prescribe how to combine them---the composition is entirely determined by the LLM.

The prompt-level constraints fall into three categories:
\begin{enumerate}
    \item \textbf{Structural constraints} (limiting the search space): at most 3 strategies per player (enforced at runtime), static single assignment (SSA), type annotations required.
    \item \textbf{Heuristic guidance} (non-restrictive): Occam's Razor (fewer strategies preferred), encouragement to break symmetry between players, penalization of duplicate algorithms.
    \item \textbf{Domain-specific constraint}: the LLM must include at least one \texttt{StationaryPoint} building block. This constraint reflects the domain knowledge that all competitive polynomial-time ANE algorithms since Tsaknakis and Spirakis~\cite{tsaknakisOptimizationApproachApproximate2007} rely on stationary-point computation; algorithms without it (e.g., KPS~\cite{kontogiannisPolynomialAlgorithmsApproximating2006}, DMP~\cite{daskalakisNoteApproximateNash2006}) achieve at best $\epsilon = 0.5$. This constraint excludes known-suboptimal directions but does not specify how the strategies returned by \texttt{StationaryPoint} should be combined---the latter is the creative part that the LLM must determine.
\end{enumerate}

Critically, the expert inputs do not include:
\begin{itemize}
    \item any suggestion of the overall algorithmic structure or composition order;
    \item any hint to use or avoid the extension technique;
    \item any specification of mixing coefficients or payoff-dependent weights;
    \item any target $\epsilon$ value;
    \item any few-shot examples of algorithm designs (the provided code examples illustrate only the expected format, not algorithmic strategies).
\end{itemize}

The results provide empirical evidence that the expert inputs did not bias the LLM toward the discovered solutions: of the 7 distinct three-player algorithms produced across 12 rounds, only the first used the extension technique; the remaining 6 independently explored non-extension structures. The final $0.5+\delta$ algorithm combines \texttt{BestResponse}, \texttt{EqMix}, and \texttt{OptimalMixing} in a pattern that was not anticipated by the human experts who designed the building blocks.

\paragraph{Process Management for Robust Exploration}
To prevent redundant exploration, we maintain a record of all previously generated algorithms. If the LLM proposes a duplicate, it is informed and shown the record of past attempts. If duplicates persist up to a certain threshold, the interaction is restarted to encourage new lines of inquiry. Similarly, to prevent the LLM from forgetting instructions in a long conversation history, we restart the interaction when the chat history exceeds a certain length. Finally, to ensure timely analysis by LegoNE, we set limits on the number of code lines and constructed strategies, prompting the LLM to simplify any overly complex algorithms it generates.

\subsection*{Experimental Setup and Main Results}

For our experiments, we employed DeepSeek-R1-250120 (abbreviated as R1)~\cite{guoDeepSeekR1IncentivizesReasoning2025}, a reasoning LLM. We set the temperature to 0.8, allowing for some variation in the LLM's responses. We provide the LLM with a set of building blocks derived from existing human-designed algorithms.

For the LegoNE analyzer, we use the same environment as in the benchmarking subsection above. For two-player cases, we set the Mathematica parameters as follows: \verb|AccuracyGoal| to 6, and \verb|MaxIterations| to 5000. For three-player cases, since the computation is more difficult, we set \verb|AccuracyGoal| to 40, \verb|WorkingPrecision| to 60, and \verb|MaxIterations| to 2000. For both cases, we use the default method (auto) as the optimization method. As a reproducibility check, we run the full process on a Windows personal computer (PC) and then manually verify the LegoNE analyzer output on the MacBook.

\paragraph{Two-Player Game Experiments}
In the two-player game setting, R1 replicated the best polynomial-time worst-case guarantee in 2 rounds, but with a different algorithm than the literature. This provides a concrete comparison with the 15 years between the previous guarantee and the current one in the human-designed literature.

We also tested with the DeepSeek-V3-241226 (abbreviated as V3) model~\cite{deepseek-aiDeepSeekV3TechnicalReport2024}. Again, we set the temperature to 0.8 and used the same building-block requirements. In a more extensive test of 100 rounds, V3 did not surpass the best polynomial-time worst-case guarantee but did discover several combinations not found in the literature.

\paragraph{Three-Player Game Experiments}
In the more challenging three-player game domain, R1 discovered an algorithm with a provable approximation guarantee of $(0.5+\delta)$ in 11 rounds. This improves upon the previously best human-designed polynomial-time worst-case guarantee of $0.6+\delta$. Furthermore, R1 discovered six additional algorithms with guarantees ranging from $0.5+\delta$ to $0.8+\delta$ within 12 rounds.

All but the first of these discovered algorithms do not use the extension technique shown in \Cref{figure:extension3p}, which was previously the only known approach for human experts to design a three-player ANE algorithm with a nontrivial guarantee better than 1. These structurally distinct algorithms show that the automated process can explore parts of the design space outside the extension paradigm. The best-performing discovered algorithm is presented in \Cref{figure:new3p}. The extension-technique constructions terminate with a best-response-then-mix step; in contrast, the discovered algorithm relies on a substantially more intricate composition of building blocks, interleaving \texttt{BestResponse}, \texttt{EqMix}, and \texttt{OptimalMixing} in a nontrivial manner.

\section*{Related Work}

\paragraph{The Computational Complexity of Nash Equilibria.}
The Nash equilibrium~\cite{nashNonCooperativeGames1951,nashEquilibriumPointsNperson1950} is the central solution concept in non-cooperative game theory. Nash's existence theorem is purely existential; the first algorithmic approach, the Lemke--Howson algorithm~\cite{lemkeEquilibriumPointsBimatrix1964}, requires exponential time in the worst case, and for decades no polynomial-time algorithm was found, though no hardness proof was known either. In this setting, Lipton, Markakis, and Mehta~\cite{liptonPlayingLargeGames2003} initiated the study of approximate Nash equilibria by proving that $\epsilon$-ANE can be computed in quasi-polynomial time for any constant $\epsilon$, via a structural result showing that Nash equilibria can be well approximated by strategies with logarithmic support---the first sub-exponential algorithm for the problem. The subsequent PPAD-completeness results of Daskalakis, Goldberg, and Papadimitriou~\cite{daskalakisComplexityComputingNash2006} (for three or more players) and Chen and Deng~\cite{chenSettlingComplexityComputing2009} (for two players) confirmed that computing an exact Nash equilibrium is intractable, providing theoretical justification for the pursuit of approximation.

Following PPAD-completeness, polynomial-time approximation algorithms advanced rapidly: Kontogiannis, Panagopoulou, and Spirakis~\cite{kontogiannisPolynomialAlgorithmsApproximating2006} achieved $\epsilon = 0.75$; Daskalakis, Mehta, and Papadimitriou~\cite{daskalakisNoteApproximateNash2006,daskalakisProgressApproximateNash2007} improved this to $0.5$ and $0.382$; Bosse, Byrka, and Markakis~\cite{bosseNewAlgorithmsApproximate2010} reached $0.364$; and Tsaknakis and Spirakis~\cite{tsaknakisOptimizationApproachApproximate2007} attained $0.3393+\delta$. Progress then stalled for 15 years. In the quasi-polynomial regime, Babichenko, Barman, and Peretz~\cite{babichenkoSimpleApproximateEquilibria2014} substantially improved the support-size bounds for multi-player games, reducing the dependence on the number of players from polynomial to logarithmic. Rubinstein~\cite{rubinsteinSettlingComplexityComputing2016} proved that no PTAS exists (assuming the Exponential Time Hypothesis for PPAD), ruling out polynomial-time algorithms for arbitrarily small $\epsilon$; however, the corresponding hardness constant is not explicit and is believed to be very small~\cite{deligkasPureCircuitTightInapproximability2024}. Deligkas, Fasoulakis, and Markakis~\cite{deligkasPolynomialtimeAlgorithm12023} finally broke the 15-year barrier with a $1/3+\delta$ guarantee.

For multi-player games, the extension technique~\cite{bosseNewAlgorithmsApproximate2010,hemonApproximateNashEquilibria2008} lifts an $r$-player algorithm to $(r{+}1)$ players with guarantee degrading from $\epsilon_r$ to $1/(2-\epsilon_r)$. Applied to the best two-player result, this yields $0.6+\delta$ for three-player games, with guarantees approaching~$1$ as the number of players grows. No alternative design paradigm for polynomial-time multi-player ANE algorithms was known prior to this work.

\paragraph{Explore-and-Evaluate Pipelines for Algorithmic Discovery.}
A broad thread of work automates algorithm development by exploring a hypothesis space and evaluating candidates with a machine-computable objective, including program synthesis~\cite{jacindhaProgramSynthesisSurvey2022} and automated algorithm configuration~\cite{stutzleAutomatedAlgorithmConfiguration2021}. Recent systems use learned heuristics to navigate large spaces of low-level algorithms, e.g., reinforcement-learning agents that discover faster sorting routines~\cite{mankowitzFasterSortingAlgorithms2023} and new matrix multiplication decompositions~\cite{fawziDiscoveringFasterMatrix2022}, as well as self-learning synthesis for structured domains such as integer-sequence programs~\cite{gauthierImprovementsProgramSynthesis2023}. These efforts typically evaluate candidates by empirical metrics (runtime, operation counts, benchmark accuracy) or by correctness for fixed input sizes, whereas LegoNE targets theorem-style worst-case approximation guarantees over an infinite family of games.

\paragraph{Relationship to AI-for-Science Systems.}
LegoNE combined with an LLM shares the high-level explore-and-evaluate architecture with recent AI-for-science systems such as AlphaGeometry~\cite{trinhSolvingOlympiadGeometry2024,chervonyiGoldmedalistPerformanceSolving2025}, FunSearch~\cite{romera-paredesMathematicalDiscoveriesProgram2024}, and AlphaEvolve~\cite{novikovAlphaEvolveCodingAgent2025}. Within this shared paradigm, three technical differences are worth noting. First, the evaluators in prior systems are either pre-existing (AlphaGeometry's deductive engine DDAR dates to the work of Chou, Gao, and Zhang in the 1990s~\cite{chouMachineProofsGeometry1994}) or straightforward to construct (FunSearch evaluates programs by executing them on specific instances); the LegoNE analyzer---based on the instantiation and forgetting compilation principles---is a formal system constructed in this work. Second, while AlphaGeometry's evaluator provides binary feedback (proof found or not), the LegoNE analyzer returns a quantitative worst-case guarantee $\epsilon$, enabling gradient-style optimization in the LLM search loop. Third, AlphaGeometry verifies static geometric propositions, whereas LegoNE verifies universal performance guarantees of procedural algorithms, requiring Floyd--Hoare-style program reasoning combined with the instantiation and forgetting compilation.

\paragraph{Sample-Based Guarantees for Strategic Properties.}
There is a growing literature that evaluates algorithms or strategic properties from samples, providing distribution-dependent guarantees rather than worst-case ones. For example, generalization results quantify how much data suffices to learn or tune high-performing algorithms under an unknown instance distribution~\cite{balcanHowMuchData2024a}, and related techniques estimate approximate incentive compatibility or distance from equilibrium in auctions from sampled types or play trajectories~\cite{balcanEstimatingApproximateIncentive2023,pierothVerifyingApproximateEquilibrium2024}. These tools are well-suited to settings where a reference distribution is meaningful and data are available, and they complement classical worst-case analyses that seek instance-independent approximation guarantees.

\paragraph{LLMs for Game Solving and Strategy/Policy Generation.}
Recent work studies LLMs as generators of strategies and policies, including self-play with code-generating models to refine game strategies~\cite{bachrachCombiningCodeGenerating2025}, guiding language generation with game-theoretic equilibrium solvers~\cite{gempSteeringLanguageModels2024}, translating natural-language game descriptions into extensive-form representations~\cite{dengNaturalLanguageExtensiveForm2025}, and code-space response oracles for interpretable multi-agent policies~\cite{hennesCodeSpaceResponseOracles2026}. These lines primarily target empirical behavior in particular games or interaction domains, whereas our focus is synthesizing polynomial-time algorithms for normal-form games together with machine-checkable worst-case guarantees.

\section*{Data Availability}
All data generated in this study---including the discovered ANE algorithms expressed in the LegoNE language, the benchmark approximation guarantees computed by the analyzer (Tables~\ref{tab:benchmark},~\ref{tab:benchmark-detail}), the LLM prompts (\verb|src/auto_design/prompts.py|), and the running-time measurements---are publicly available at the Zenodo repository~\cite{liLegoNEZenodo2026}; reuse of these materials is governed by the license described in the Code Availability statement below.

\section*{Code Availability}

The source code of the LegoNE compiler and analyzer, the building-block library, and the LLM-driven discovery pipeline used in this study is deposited in the Zenodo repository~\cite{liLegoNEZenodo2026}. The code includes implementations sufficient to reproduce all experiments reported in this paper. The repository is released under copyright ``All rights reserved'' by the authors, owing to ongoing patent-related plans. Non-commercial academic use---including reproduction, verification, and extension of the results reported in this paper---is permitted free of charge upon written request to the corresponding author Hanyu Li (\texttt{lhydave@pku.edu.cn}); requests are typically answered within two weeks. The repository will remain available indefinitely. Other uses require a separate licensing agreement with the authors.

\section*{Acknowledgements}
The authors would like to thank Ruyi Ji, Yuhao Li, and Paul Spirakis for helpful discussions. The authors would also like to thank the anonymous reviewers for their valuable comments and suggestions.

\section*{Funding}
This work is supported by the Natural Science Foundation of China (Grant No.~62572010 and Grant No.~6212290003).

\section*{Author Contributions}
H.L.\ and D.L.\ conceived the project and designed the LegoNE framework. H.L.\ implemented the LLM-powered discovery loop and the LegoNE building-block library, and carried out the computational experiments. D.L.\ contributed to experimental analysis and validation. H.L.\ and D.L.\ jointly developed the formal analyses and proofs. H.L.\ wrote the original draft; H.L.\ and D.L.\ contributed to the writing and led the revision responses. X.D.\ supervised the project.

\section*{Competing Interests}
The authors declare no competing interests.

\clearpage
\bibliographystyle{naturemagdoi}
\bibliography{AutoAI,ApproxNE}

\clearpage
\appendix
\section{A Formal and Complete Treatment of LegoNE}\label{sec:formal-treatment}

In this part, we provide a formal and complete treatment of LegoNE. We first define the syntax of logic encodings of building blocks, algorithms, and the proof goal. Then, we describe the instantiation and forgetting tactics. Finally, we show how to convert the LegoNE code into a constrained optimization problem to find the approximation guarantee.

\subsection{Syntax of Logic Encodings}\label{subsec:syntax-logic-encodings}

We first define the syntax of logic encodings of building blocks.

\paragraph*{Symbols.}

We use the following symbols in the logic encodings:
\begin{itemize}
    \item payoff functions: $u_1, u_2, \dots, u_r$,
    \item loss functions: $f_1, f_2, \dots, f_r,f$,
    \item strategy variables: $\x^1, \x^2, \dots, \x^r, \y^1, \y^2, \dots, \y^r,\dots$,
    \item payoff variables: $U_1, U_2, \dots$,
    \item real variables: $\rho_1, \rho_2, \dots$,
    \item real constants: $1, -1, 0.5, \dots$,
    \item arithmetic operators: $+,-,\times,\div,\leq,\geq,=,\min,\max$,
    \item logic connectives: $\wedge, \vee$,
    \item quantifiers: $\forall, \exists$.
    \item other symbols: $(,),:,,$.
\end{itemize}

\paragraph*{Terms.}

We can only use the following terms in the logic encodings:
\begin{itemize}
    \item payoff terms: $U(\x^i, \y^{-i})$, where $U$ is among payoff functions $u_1, u_2, \dots, u_r$ and payoff variables $U_1, U_2, \dots$,
    \item loss terms: $f_i(\x^i, \y^{-i})$,
    \item maximum of payoff terms: $\max_{\x^i} U(\x^i, \y^{-i})$,
    \item real variables: $\rho_i$,
    \item real constants: $1, -1, 0.5, \dots$.
\end{itemize}

\paragraph*{Arithmetic expressions.} 

The set $L_A$ of arithmetic expressions can be inductively defined as follows:
\begin{itemize}
    \item Terms are in $L_A$.
    \item If $\alpha, \beta\in L_A$, then $(\alpha + \beta), (\alpha - \beta), (\alpha \times \beta), (\alpha \div \beta), \min\{\alpha, \beta\}, \max\{\alpha, \beta\}\in L_A$. 
\end{itemize}

\begin{remark}
    Note that we set the maximum of payoff terms as a term, rather than an expression. This is because the maximum operation here ranges over $\x^i$, which is not a finite set. Thus, the arithmetic properties of such maximum operations are not straightforward, compared to the maximum of two real numbers. Humans need to provide the properties of such maximum operations in the logic encodings.

    Similarly, while we know that payoff functions are multi-linear, it could be difficult for machines to flexibly utilize multi-linearity. Thus, we only treat payoff functions as terms, rather than expressions. Multi-linearity is then provided by humans as logic encodings, which are much easier for machines to utilize.
\end{remark}

\paragraph*{Comparison expressions.}

A comparison expression is a formula in the form of $\alpha ~ \mathsf{COMP} ~ \beta$, where $\alpha, \beta\in L_A$ and $\mathsf{COMP}\in\{\leq, \geq, =\}$.

\paragraph*{Basic types.}

We use $a:A$ to denote that $a$ is of type $A$. We have the following basic types and their elements used to define the input and output of building blocks:
\begin{itemize}
    \item None type $\mathsf{None}$: no element,
    \item Real type $\mathsf{Real}$: arithmetic expressions in $L_A$,
    \item Payoff type $\mathsf{Payoff}$: payoff functions $u_1, u_2, \dots, u_r$ and their linear combinations with coefficients in $\mathsf{Real}$,
    \item Comparison type $\mathsf{Comp}$: comparison expressions,
    \item Player $i$'s strategy type $\mathsf{Strategy}_i$: the element in $\mathsf{Strategy}_i$ can be inductively constructed as follows. If $\x^i, \y^i: \mathsf{Strategy}_i$, then $\alpha \x^i + (1-\alpha)\y^i$ is in $\mathsf{Strategy}_i$ for any $\alpha:\mathsf{Real}$ with $\alpha\in[0,1]$.
\end{itemize}

\paragraph*{Building blocks.}

A building block is a function with inputs and outputs. The type of building blocks is
\begin{equation}
T_{i1}\times T_{i2}\times\dots\times T_{ik}\to T_{o1}\times T_{o2}\times\dots\times T_{om},
\end{equation}
where $T_{ij}$'s are input types and $T_{ok}$'s are output types. The input types can be any of the basic types above, and the output types can be any of them except the comparison type. For example, when $r=3$, $\mathsf{BestResponse2}$ is a building block with type $\mathsf{Strategy}_1\times \mathsf{Strategy}_3\to \mathsf{Strategy}_2$.

\paragraph*{Logic encodings of building blocks.}

We use $\phi$ to denote the mapping from a building block to a logic encoding. A building block can be encoded as a formula from the following set $L_E$, inductively constructed as follows:
\begin{itemize}
    \item All atomic properties $\alpha$'s belong to $L_E$, where each $\alpha$ has the following form:
    \begin{equation}
        (\exists\rho_1\dots,\rho_s)(\forall \x_1^1, \dots, \x_{m_r}^r)(\forall U_1, U_2, \dots)\gamma.\label{eq:general-encoding}
    \end{equation}
    The meaning of each symbol is as follows:
    \begin{itemize}
        \item $\rho_i$ is a real variable.
        \item $\x^i_t$ is the $t$-th strategy variable of player $i$.
        \item $U_i$ is a payoff variable.
        \item $\gamma$ is a comparison expression.
    \end{itemize}
    \item If $\alpha,\beta\in L_E$, then $(\alpha\wedge\beta)\in L_E$.
    \item If $\alpha,\beta\in L_E$, then $(\alpha\vee\beta)\in L_E$.
\end{itemize}

Logic encodings are used to describe the properties of building blocks. For example, in two-player games, the building block $\y^1=\mathsf{BestResponse1}(\x^2)$ can be encoded as
\begin{equation}
\forall \z^1(u_1(\z^1, \x^2)\leq u_1(\y^1, \x^2)).
\end{equation}
This encoding only have an atomic property. 

We can have more complicated properties by using conjunctions and disjunctions of such atomic properties. The conjunction can be understood as ``this building block has all these properties'', while the disjunction can be understood as ``this building block has at least one of these properties''. Finally, the existential quantifiers can be understood as ``this building block constructs some internal variables to achieve these properties''. For example, the encoding of building block $\mathsf{StationaryPoint}$ has an existential quantifier $\exists \rho$. (See \Cref{sec-legoNE-constraint-embedding})

Apart from the building blocks, the definitions of payoff function $u_i$ and regret $f_i$ themselves also imply a group of \emph{inherent formulas}:

\begin{itemize}
    \item For $i\in [r]$, $\forall\x(0\leq u_i(\x)\leq 1\wedge 0\leq f_i(\x)\leq 1)$.
    \item For $i\in [r]$, $\forall\x^i,\forall\x^{-i}(f_i(\x^i,\x^{-i})=\max_{\tilde{\x}^i} u_i(\tilde{\x}^i,\x^{-i})-u_i(\x^i,\x^{-i}))$.
    \item For $i\in [r]$, $\forall\x^i,\forall\x^{-i}(\max_{\tilde{\x}^i} u_i(\tilde{\x}^i,\x^{-i})\geq u_i(\x^i,\x^{-i}))$.
    \item For $i\in [r]$, $\forall\x^{-i}(\max_{\tilde{\x}^i} u_i(\tilde{\x}^i,\x^{-i})\leq 1)$.
\end{itemize}
We denote the conjunction of these formulas for all $i$'s as $\phi_0$.

\paragraph*{Algorithms.}

In the LegoNE framework, each algorithm is a sequence of assignment statements. One assignment uses a defined building block with existing variables as inputs and new variables as outputs. 

Formally, suppose we have defined a set $\mathcal{F}$ of building blocks. Then, an algorithm is a finite sequence of assignments $s_1,\dots,s_p$. For each assignment $s_i$, we maintain a corresponding variable set $V_i$. $V_0$ is the input of the algorithm. $s_i$ must be in one of the following forms:
\begin{enumerate}
    \item $v'_1,\dots,v'_m= f_i(v_1,\dots,v_n)$, where $v'_1,\dots,v'_m$ are new variables not in $V_{i-1}$, $v_1,\dots,v_n$ are existing variables in $V_{i-1}$ or constants, and $f_i$ is a building block in $\mathcal{F}$.
    \item return $v_1,\dots,v_r$, where $v_1,\dots,v_r$ are existing variables in  $V_{i-1}$ and variable $v_i$ has type $\mathsf{Strategy}_i$.
\end{enumerate}
If form 1 is used, then update $V_i=V_{i-1}\cup \{v'_1,\dots,v'_m\}$; otherwise $V_i=V_{i-1}$.

A more intuitive way to write an algorithm is given as follows:
\begin{algorithm}[H]
    \caption{Algorithm $\Gamma$}
    \begin{algorithmic}[1]
    \Require Payoff functions $u_1, u_2, \dots, u_r$
    
    \lineno{1} $s_{11},s_{12},\dots,s_{1m_1}=B_1$
    
    \lineno{2} $s_{21},s_{22},\dots,s_{2m_2}=B_2$
    
    $\dots$

    \lineno{k} $s_{k1},s_{k2},\dots,s_{km_k}=B_k$
    
    \lineno{k+1} \textbf{return} $s^1,s^2,\dots,s^r$
\end{algorithmic}
\end{algorithm}
Here, $s_{ij}$ is a strategy as the $j$-th output of building block $B_i$ using only variables $s_{uv}$ with $u<i,v\leq m_u$, and $s^i$ is player $i$'s strategy, being one among $s_{11},\dots,s_{km_k}$.

\paragraph*{Encoding of algorithms.}

The encoding of algorithm $\Gamma$ is
\begin{equation}
\phi[\Gamma]: \phi_0 \wedge \phi[\text{line $1$}] \wedge \dots \wedge \phi[\text{line $k$}].
\end{equation}

\paragraph*{Proof goal.}

To show that algorithm $\Gamma$ has an approximation guarantee of $b$, the proof goal is
\begin{equation}
    (\forall u_1, u_2, \dots, u_r) (\forall s_{11}, \dots, s_{km_k}) (\phi[\Gamma] \to f(s^1, s^2, \dots, s^r) \leq b).\label{eq:formal-proof-goal}
\end{equation}

\subsection{Instantiation and Forgetting}

In this part, we formally give the instantiation and forgetting tactics in LegoNE. These tactics are essential for rewriting the proof goal into a form that is amenable to optimization.

Here, we propose \Cref{algo:instantiation} to instantiate the universal quantifiers in any property in the form of \eqref{eq:general-encoding}.

\begin{algorithm}[htb]
    \caption{Instantiation Algorithm}\label{algo:instantiation}
    \begin{algorithmic}[1]
    \Require A logic formula $\psi$ in the form of \eqref{eq:general-encoding}, a series of payoff functions $u_1, u_2, \dots, u_r$, and a series of strategies $s^1_1, s^1_2, \dots, s^r_{m_r}$.
    \Ensure A series of inequalities derived from $\psi$.
    \State For each payoff quantifier $\forall U_i$, replace it with $u_k$ for all $k\in [r]$ and all newly constructed payoff functions. \Comment{This will eliminate all payoff quantifiers.}
    \State Do the following two procedures, respectively.
    \Statex
    \State \textbf{Procedure 1: Instantiation using existing strategies.}
    \State \indent  \parbox[t]{\dimexpr\linewidth-\algorithmicindent-\algorithmicindent}{%
    For each strategy quantifier $\forall \x^i_t$, replace all occurrence of $\x^i_t$ with $s^i_k$ for all $k\in [m_i]$ and remove the quantifier $\forall \x^i_t$. \Comment{This will eliminate all strategy quantifiers.}}
        
    \State \indent  \parbox[t]{\dimexpr\linewidth-\algorithmicindent-\algorithmicindent}{%
    The remaining formulas are all universal-quantifier-free. Let $C_1$ be the conjunction of these formulas.\\}
    \State \textbf{end Procedure 1}
    \Statex
    \State \textbf{Procedure 2: Instantiation using the $\max$ operator.}
    \For{strategy quantifier $\forall \x^i_t$}
    \If {$\x^i_t$ occurs more than once in $\psi$}
    \State  \parbox[t]{\dimexpr\linewidth-\algorithmicindent-\algorithmicindent}{%
    \textbf{continue} \Comment{The maximum of two different terms involving $\x^i_t$ may not be obtained at the same $\x^i_t$.}}
    \EndIf
    \State \parbox[t]{\dimexpr\linewidth-\algorithmicindent}{%
    Use Procedure 1 to instantiate all but one strategy quantifier $\x^i_t$ and obtain $C_1$. This will eliminate all but one strategy quantifier.}
    \State For every term $u_i(\x^i_t, \cdot)$ in $C_1$, replace it with $\max_{\x^i_t} u_i(\x^i_t, \cdot)$. The result is $C_2(\x^i_t)$.
    \EndFor
    \State Let $C_2$ be the conjunction of $C_2(\x^i_t)$ for all $\x^i_t$ (if any).
    \State \textbf{end Procedure 2}
    \Statex
    \State \Return $C_1\wedge C_2$.
    \end{algorithmic}
\end{algorithm}

Note that \Cref{algo:instantiation} only instantiates the atomic properties \eqref{eq:general-encoding} in the logic encoding. The instantiation of the whole logic encoding is given by simply instantiating all atomic properties in the encoding.

Suppose we do the instantiation for the proof goal \eqref{eq:formal-proof-goal} of algorithm $\Gamma$. We denote the result as $\psi'$. Now we describe the forgetting tactic.

For the forgetting tactic, it is essentially a renaming process. Since we have formally defined what a term is, we can directly rename all terms using new names such that different terms have different names. We present it in \Cref{algo:forgetting}.

\begin{algorithm}[htb]
    \caption{Forgetting Algorithm}\label{algo:forgetting}
    \begin{algorithmic}[1]
    \Require A logic formula $\psi'$ after instantiation of the proof goal \eqref{eq:formal-proof-goal}.
    \State Collect all terms in $\psi'$. Let $T$ be the set of all terms.
    \State For each term $t\in T$, define a new symbol $\theta(t)$ that does not appear in $\psi'$.
    \State Replace each term $t$ in $\psi'$ with $\theta(t)$.
    \State Remove all quantifiers in outermost of $\psi'$ and replace them with $\forall \theta(t)$ for all $t\in T$.
    \State \Return $\psi'$.
    \end{algorithmic}
\end{algorithm}

Using the instantiation and forgetting tactics, we can rewrite the proof goal into a form that is ready for optimization. We will discuss this in the next section.

\subsection{Converting LegoNE Code into an Optimization Problem}

In this part, we show how to convert the LegoNE code into a constrained optimization problem to find the approximation guarantee.

We still focus on the general Algorithm $\Gamma$ in \Cref{subsec:syntax-logic-encodings}. This time, we do not know the approximation guarantee of the algorithm. Instead, we set an unknown approximation guarantee $b$, but the proof goal \eqref{eq:formal-proof-goal} remains the same. For convenience, we here write down the proof goal again:
\begin{equation}(\forall u_1, u_2, \dots, u_r) (\forall s_{11}, \dots, s_{km_k}) (\phi[\Gamma] \to f(s^1, s^2, \dots, s^r) \leq b).\end{equation}
We need to find a real number $b$ as small as possible such that the above formula is valid.

Similarly, we can rewrite the proof goal into
\begin{equation}
\begin{aligned}
    \forall a_1,\dots,\forall b_1,\dots,\forall c_1,\dots(\phi'[\Gamma]\to g(a_1,\dots,b_1,\dots,c_1,\dots)\leq b),
\end{aligned}\label{eq:unknown-eliminated}
\end{equation}

Here, $\phi'[\Gamma]$ contains no universal quantifiers on payoffs or strategies. However, in the general case, $\phi'[\Gamma]$ still contains existential quantifiers on real variables. We need to handle these existential quantifiers.

The existential quantifier can be handled by the following standard result in predicate logic (see. e.g., Theorem 3.5.11 on page 74 and Example 1 on page 73 in textbook \cite{vandalenLogicStructure2013}).

\begin{lemma}\label{lemma:legoNE-existential-quantifier}
    Suppose $x$ occurs free in formula $A(x)$ and does not occur free in formula $B$. Then 
\begin{itemize}
    \item $(\exists x A(x))\to B$ is logically equivalent to $\forall x (A(x)\to B)$.
    \item $(\exists x A(x))\wedge B$ is logically equivalent to $\exists x (A(x)\wedge B)$.
    \item $(\exists x A(x))\vee B$ is logically equivalent to $\exists x (A(x)\vee B)$.
\end{itemize}
\end{lemma}
Using the second and the third properties, we can rewrite the proof goal \eqref{eq:unknown-eliminated} as 
\begin{equation}
    \begin{aligned}
        \forall a_1,\dots,\forall b_1,\dots,\forall c_1,\dots,\forall \rho_1\dots(\phi''[\Gamma]\to g(a_1,\dots,b_1,\dots,c_1,\dots)\leq b),
    \end{aligned}\label{eq:unknown-eliminated-existential}
\end{equation}
where $\phi''[\Gamma]$ is quantifier-free.

    

To find the smallest $b$ such that \eqref{eq:unknown-eliminated-existential} is valid, we can instead find the largest $g$ over all possible values of $a_1,\dots,b_1,\dots,c_1,\dots,\rho_1,\dots$ and set $b$ to be this largest value. Thus, we can write down the following constrained optimization problem:
\begin{equation}
\begin{aligned}
        \operatorname*{maximize}_{a_1,\dots,b_1,\dots,c_1,\dots,\rho_1,\dots} &\quad g(a_1,\dots,b_1,\dots,c_1,\dots)\\
        \text{subject to} &\quad\phi''[\Gamma].
\end{aligned}\label{eq:legoNE-optimization-problem}
\end{equation}

Then we show how to solve this optimization problem. 

First, since constraint $\phi''[\Gamma]$ is quantifier-free, we can write the premise into disjunctive normal form (DNF). This step can also be performed by machines \cite{vandalenLogicStructure2013}. An exemplary form of DNF for $\phi''[\Gamma]$ is $(a_1\leq a_2\wedge a_5\geq a_6)\vee(b_1=b_2\wedge b_3\geq b_4)\vee(c_1\geq c_2)$.

Second, \eqref{eq:legoNE-optimization-problem} can be split into three optimization problems: to maximize $g$ subject to $a_1\leq a_2\wedge a_5\geq a_6$, to maximize $g$ subject to $b_1=b_2\wedge b_3\geq b_4$, and to maximize $g$ subject to $c_1\geq c_2$. Each of these is a fixed-size constrained optimization problem, and thus can be solved by numerical solvers (like Mathematica \cite{WolframMathematicaModern} or Gurobi \cite{GurobiOptimization}). Suppose the optimal values of these problems are $v_1,v_2,v_3$. Then, we actually prove that
\begin{itemize}
    \item if $b_1=b_2\wedge b_3\geq b_4$, then $g(\dots)\leq v_1$;
    \item if $a_1\leq a_2\wedge a_5\geq a_6$, then $g(\dots)\leq v_2$;
    \item if $c_1\geq c_2$, then $g(\dots)\leq v_3$.
\end{itemize}
Finally, combining these results, we show that under constraint $\phi''[\Gamma]$, $g(\dots)\leq\max\{v_1,v_2,v_3\}$. Thus, we can set $b=\max\{v_1,v_2,v_3\}$ as the final approximation guarantee.

\clearpage
\section{Building blocks and their logic encodings}\label{sec-legoNE-constraint-embedding}

In this part, we show the building blocks in the literature and their logic encodings. For an accessible presentation, we will use mathematical symbols rather than the LegoNE code. It is not hard to see the equivalence between the mathematical symbols and the LegoNE code. 

\subsection{Inherent constraints}

The definitions of payoff functions and regret functions themselves provide constraints. Here, we list the inherent constraints for the payoff functions and regret functions.

\begin{itemize}
    \item For given $i\in [r]$, $\forall\x(0\leq u_i(\x)\leq 1)$.
    \item For given $i\in [r]$, $\forall\x^i,\forall\x^{-i}(f_i(\x^i,\x^{-i})=\max_{\x^{i'}} u_i(\x^{i'},\x^{-i})-u_i(\x^i,\x^{-i}))$.
    \item For given $i\in [r]$, $\forall\x^i,\forall\x^{-i}(\max_{\x^{i'}} u_i(\x^{i'},\x^{-i})\geq u_i(\x^i,\x^{-i}))$.
    \item For given $i\in [r]$, $\forall\x^{-i}(\max_{\x^{i'}} u_i(\x^{i'},\x^{-i})\leq 1)$.
\end{itemize}

\subsection{Building blocks in the literature}\label{sec-legoNE-basic-operations}

We can collect most basic (polynomial-time computable) operations from the literature. Although most of them are designed for two-player games, we can easily generalize them under the setting of $r$-player games. 
\begin{itemize}
    \item \textbf{Random strategy}: $\x^i= \mathsf{Randomi}()$.
    
    \emph{Description}: Sample a random strategy $\x^i$ for player $i$.

    \emph{Logic encoding}: None.

    \item \textbf{Best response} (first occurred in \cite{daskalakisNoteApproximateNash2006}): $\x^i= \mathsf{BestResponse}(\x^{-i})$.
    
    \emph{Description}: For a player $i$, given a strategy profile $\x^{-i}$ of the other players, find the best response strategy $\x^i=\arg \max_{\x^i\in\Delta_{n_i}} u_i(\x^i,\x^{-i})$.
    
    \emph{Logic encoding}: 
    \begin{equation}\forall \x^{i'}(u_i(\x^i, \x^{-i})\geq u_i(\x^{i'}, \x^{-i})).\end{equation} 
    \begin{remark}
        The operation described in mathematical symbols is a ``template'' for the real code. In the LegoNE code, we need to instruct which $i$ is used in the operation. Thus, the real code should be \verb|BestResponsei| for player $i$.
    \end{remark}

    \item \textbf{Zero-sum NE} (first occurred in \cite{bosseNewAlgorithmsApproximate2010}): $\x^i,\x^j= \mathsf{ZeroSumNE}(\x^{-i,j},u)$.
    
    \emph{Description}: For a given linear combination $u$ of payoffs $u_1,\dots,u_r$ and a strategy profile $\x^{-{i,j}}$ of the other players, find the NE $(\x^i,\x^j)$ of zero-sum two-player game $u(\cdot,\cdot, \x^{-i,j})$.

    \emph{Logic encoding}:
    \begin{equation}
        \begin{aligned}
        \forall \x^{i'}(u(\x^{i'},\x^j,\x^{-i,j}) \geq u(\x^i,\x^j,\x^{-i,j}))~\wedge\\
        \forall \x^{j'}(u(\x^{i},\x^{j'},\x^{-i,j}) \leq u(\x^i,\x^j,\x^{-i,j})).
        \end{aligned}
    \end{equation}
	    \item \textbf{Stationary point} (first occurred in \cite{tsaknakisOptimizationApproachApproximate2007}): $\x^i,\x^j,\y^i, \y^j= \mathsf{StationaryPoint}(\x^{-i,j})$.

		    \emph{Description}: Given a strategy profile $\x^{-{i,j}}$ of the other players, \emph{hold it fixed} and consider the two-variable objective $\max\{f_i(\x^i,\x^j,\x^{-i,j}),\,f_j(\x^i,\x^j,\x^{-i,j})\}$ over players $i$ and $j$'s mixed strategies. The building block returns a pair $(\x^i,\x^j)$ that is a stationary point of this objective using linear programming, associated with two dual solutions $\y^i$ and $\y^j$.
	    
		    \begin{remark}
		        In the main text, \texttt{StationaryPoint3} is used for three-player games. It is simply the above \texttt{StationaryPoint} instantiated with players $(i,j)=(1,2)$ and $\x^{-1,2}=z$, where $z$ is player 3's current mixed strategy; it outputs $(x_s,y_s,w,z_{\mathrm{dual}})$ corresponding to $(\x^1,\x^2,\y^1,\y^2)$.
			    \end{remark}

	    \emph{Logic encoding}:
    \begin{equation}
        \begin{gathered}
            f_i(\x^i,\x^j,\x^{-i,j})=f_j(\x^i,\x^j,\x^{-i,j})~\wedge\\
            \forall \x^{i'}(u_i(\y^i,\x^{j},\x^{-i,j}) \geq u_i(\x^{i'},\x^j,\x^{-i,j}))~\wedge\\
            \forall \x^{j'}(u_j(\x^{i},\y^j,\x^{-i,j}) \geq u_j(\x^i,\x^{j'},\x^{-i,j}))~\wedge\\
            \exists \rho\in [0,1], \forall \x^{i'}, \forall \x^{j'},\\
            f_i(\x^i,\x^j,\x^{-i,j})\leq \rho \left( u_i(\y^i, \x^{j'}, \x^{-i,j})-u_i(\x^{i'}, \x^j, \x^{-i,j})-u_i(\x^i, \x^{j'}, \x^{-i,j})+u_i(\x^i, \x^j, \x^{-i,j}) \right)+\\
            (1-\rho) \left( u_j(\x^{i'}, \y^j, \x^{-i,j})-u_j(\x^{i'}, \x^j, \x^{-i,j})-u_j(\x^i, \x^{j'}, \x^{-i,j})+u_j(\x^i, \x^j, \x^{-i,j}) \right).
        \end{gathered}  
    \end{equation}
    Note that the quantifier $\exists\rho\in[0,1]$ is treated as a parameter in the LegoNE code.

    \item \textbf{Uniform mixing} (first occurred in \cite{kontogiannisPolynomialAlgorithmsApproximating2006}): $\x^i= \mathsf{UniformMixing}(\x^{i}_1,\dots,\x^{i}_s)$
    
    \emph{Description}: For a player $i$, given strategies $\x^{i}_1,\dots,\x^{i}_s$, output the strategy $\x^i=\frac{1}{s}\sum_{k=1}^s \x^{i}_k$.

    \emph{Logic encoding}: 
    \begin{equation}\begin{gathered}
    \forall u\in\text{Payoff},\forall \x^{-i}\left(u(\x^i, \x^{-i})=\frac{1}{s}\sum_{k=1}^s u(\x^{i_k}, \x^{-i})\right)\wedge\\
    \forall x^{-i}\left(\sum_{k=1}^s \frac{1}{s}f_j(\x^{i_k}, \x^{-i})\geq f_j(\x^i, \x^{-i})\right)~\text{for all } j\in [r], j\neq i.
    \end{gathered}\end{equation}
    The second part of the formula is the Jensen's inequality over the convex function $f_j$.

    \item \textbf{Branching}: $\mathsf{IfThenElse}(a,b,\text{branch 1}, \text{branch 2})$
    
    \emph{Description}: Given certain values $a$, $b$, go to branch $1$ if $a\geq b$, and go to branch $2$ otherwise.

   \emph{Logic encoding}: 
   
   \begin{equation}[(a\geq b)\wedge \text{constraints in branch 1}] \vee [(a< b)\wedge \text{constraints in branch }2].\end{equation}
    Note that operations $\max$ and $\min$ can be expressed by branch. 

    \item \textbf{Optimal mixing} (occurred in every paper on ANE algorithms, first formalized in \cite{dengOptimalMixingProblem2025}):
    \begin{equation}
    (\s^1_*,\dots,\s^r_*)= \mathsf{OptimalMixing}(\s^1_1,\dots,\s^1_{t_1},\dots,\s^r_{1},\dots,\s^r_{t_r}, u_1, \dots, u_r).
    \end{equation}

    \emph{Description}: Given a set of strategies $\s^1_1,\dots,\s^1_{t_1}\in\Delta_{n_1},\dots,\s^r_{1},\dots,\s^r_{t_r}\in\Delta_{n_r}$, and payoff functions $u_1,\dots,u_r$, the optimal mixing operation $\mathsf{OptimalMixing}$ outputs the strategy profile $(\s^1_*,\dots,\s^r_*)$ that minimizes the function $f$ on $\mathcal{M}$, the set of all convex combinations of the strategies, i.e., 
    \begin{equation}\begin{aligned}
        \mathcal{M}=\left\{(\s^1,\dots,\s^r)\in\Delta_{n_1}\times\dots\times\Delta_{n_r}:\s^k=\sum_{i=1}^{t_k}\balpha^k_i\s^{k}_i;\balpha^k_j\geq 0,j\in[t_k];\sum_{i=1}^{t_k}\balpha^k_i=1,k\in[r]\right\}.
    \end{aligned}\end{equation}

    \emph{Logic encoding}: It is very complicated and non-trivial to write down the logic encoding of the optimal mixing operation. We defer it to \Cref{subsec:encoding-optmix}.
\end{itemize}

These operations are the basic building blocks for writing polynomial-time ANE algorithms.

\subsection{Encoding of Optimal Mixing}\label{subsec:encoding-optmix}

The optimal mixing operation is the solution to a continuous optimization problem, which is not directly expressible in logic formulas. Instead, we provide a formula that is \emph{implied} by the optimal mixing operation, namely, a necessary condition of the optimal mixing operation. Formally, we have the following theorem.

\begin{theorem}\label{thm:optimal-mixing-upper-bound}
    For any fixed $r$ and $t_1, \dots, t_r$, if $\x^1_o, \dots, \x^r_o$ are the output of the optimal mixing operation given the input strategies $\s^k_i$, then there is a term $L^*$ such that 
    \begin{equation}\phi\big[(\x_o^1, \dots, \x_o^r) = \mathsf{OptimalMixing}(\s^1_1, \dots, \s^r_{t_r})\big] \mapsto [f(\x^1_o, \dots, \x^r_o) \leq L^*]\end{equation}
    and $L^*$ is expressed by $f_i$ values on the input strategies $\s^k_i$ with arithmetic operations $+,-,\times,\div,<,>$, $\max, \min$ operations over finite elements, and branch operations.
\end{theorem}

Below we first give a sketch of the proof of \Cref{thm:optimal-mixing-upper-bound}. An overall tactic is to gradually relax $f(\x^1_o,\dots,\x^r_o)$ to obtain the upper bound in the theorem. 
\begin{itemize}
    \item First, we restrict the domain from $\mathcal{M}$ to the edges of $\mathcal{M}$, which produces the first upper bound.
    \item Then, along each edge, replace $f_i$ with a linear upper bound $l_i$. The minimum of $f$ on an edge is bounded by the minimum of $\max\{l_1,\dots,l_r\}$, which can be explicitly expressed by the $f_i$ values on the vertices of the edge.
    \item Finally, the vertices of $\mathcal{M}$ are exactly the strategy profile constructed by the input. Thus, we can find an upper bound that is expressed by $f_i$ values on the constructed strategies.
\end{itemize}

Now we provide the detailed proof of \Cref{thm:optimal-mixing-upper-bound}. To simplify the notations, we below assume that the number of input strategies for each player is the same, i.e., $t_1=\dots=t_r=t$, and the action sets are of the same size, i.e., $n_1=\dots=n_r=n$. One can easily see that the general case follows the same procedure.

To further simplify the notations, we will use coefficients rather than strategies to represent the elements in $\mathcal{M}$. Note that each element in $\mathcal{M}$ can be represented by its coefficients $(\balpha^k_i)_{k,i}$. The coefficients form a set $\mathcal{A}=\Delta_{t}^r$. The correspondence between the coefficients and the strategies is given by the following diagram:

\begin{equation}
\begin{tikzcd}
    \balpha^k=(\balpha^k_1,\dots,\balpha^k_t)\in\Delta_t \arrow[r, leftrightarrow] & \sum_{i=1}^t\balpha^k_i\s^k_i\in\Delta_n\\
    \balpha=(\balpha^1,\dots,\balpha^r)\in\mathcal{A} \arrow[r, leftrightarrow] & \left(\sum_{i=1}^t\balpha^1_i\s^1_i,\dots,\sum_{i=1}^t\balpha^r_i\s^r_i\right)\in\mathcal{M}
\end{tikzcd}
\end{equation}

Then, we naturally define $f_i(\balpha)$ and $f(\balpha)$ by its corresponding strategy profiles. Now, we can give the detailed logic encoding of the optimal mixing operation.

First, $\mathcal{A}$ is a polytope, whose vertices are in form of $(\balpha^1,\dots,\balpha^r)$, where $\balpha^k\in\{\e_1,\dots,\e_t\}$. The vertices correspond to the strategy profiles constructed by the inputs $\s^k_j$. The edges of $\mathcal{A}$ are in form of
\begin{equation}
    \left\{(\e_{i_1},\dots,\e_{i_{k-1}},\lambda \e_{i_k}+(1-\lambda)\e_{i_{k}'},\e_{i_{k+1}},\dots,\e_{i_r}):\lambda\in[0,1],i_1,\dots,i_r\in[t]\right\}.
\end{equation}

Intuitively, an edge is the segment between two adjacent vertices, with only one $\balpha^k$ varying. To simplify the discussion, when we refer to a specific edge $E$, we will use the above form to represent it. Denote the union of all edges as $\mathcal{E}$. Then, we clearly have the following lemma.

\begin{lemma}\label{lemma:optimal-mixing-upper-bound}
    $\mathcal{E}\subseteq\mathcal{A}$. Thus, 
\begin{equation}
    \min_{\balpha\in\mathcal{A}}f(\balpha)\leq\min_{\balpha\in\mathcal{E}}f(\balpha)=\min_{E\text{ is an edge of }\mathcal{A}}\min_{\balpha\in E}f(\balpha).\label{eq:whole-to-edge}
\end{equation}
\end{lemma}

By this lemma, if we can find an upper bound of $\min_{\balpha\in E}f(\balpha)$ for each edge $E$, then we can find an upper bound of $\min_{\balpha\in\mathcal{A}}f(\balpha)$.

An important observation of $f_i$ is that along each edge $E$, only one $\balpha^k=\lambda \e_{i_k}+(1-\lambda)\e_{i_{k}'}$ varies, and thus $f_i$ is a function in $\lambda$. By scrutinizing the definition of $f_i$, we have the following lemma.

\begin{lemma}\label{lemma:f-i-convex-on-edge}
    For each edge $E$ and each $i\in[r]$, $f_i$ is a convex function on $E$.
\end{lemma}

\begin{proof}
    Then, by the definition of $f_i$, $f_i(\balpha)=g(\lambda)$, where $g$ has the form 
    \begin{equation}\max\{\ba_1^\T(\lambda \e_{i_k}+(1-\lambda)\e_{i_k'}),\dots,\ba_n^\T(\lambda \e_{i_k}+(1-\lambda)\e_{i_k'})\}-\bb^\T(\lambda \e_{i_k}+(1-\lambda)\e_{i_k'}).\end{equation}
    Since $\max$ is a convex function and the addition/composition of convex function with a linear function is still convex, $g$ is convex. Thus, $f_i$ is convex on $E$.
\end{proof}

By \Cref{lemma:f-i-convex-on-edge}, using Jensen's inequality, we have the following lemma, which actually gives an upper bound of $f_i$ on each edge.

\begin{lemma}\label{lemma:f-i-upper-bound}
    For each edge $E$ and each $i\in[r]$, let $\balpha^k=\lambda \e_{i_k}+(1-\lambda)\e_{i_{k}'}$ be the varying coefficient on $E$. Then, $f_i(\balpha)\leq \lambda f_i(\e_{i_1}, \dots, \e_{i_k},\dots,\e_{i_r})+(1-\lambda)f_i(\e_{i_1},\dots,\e_{i_{k}'},\dots,\e_{i_r})$.
\end{lemma}

Now, the upper bound given by \eqref{eq:whole-to-edge} becomes:

\begin{equation}
    \min_{\balpha\in\mathcal{A}}f(\balpha)\leq\min_{E\text{ is an edge of }\mathcal{A}}\underbrace{\min_{\lambda \in[0,1]}\max_{i\in[r]}\left\{\lambda f_i(\dots,\e_{i_k},\dots)+(1-\lambda)f_i(\dots,\e_{i_{k}'},\dots)\right\}}_{T_E}.
\end{equation}

Finally, to prove \Cref{thm:optimal-mixing-upper-bound}, we need to show that $T_E$ can be explicitly expressed by $f_i$ values on the vertices of $E$. The explicit expression of $T_E$ can be given by the following lemma.

\begin{lemma}\label{lemma:T-E-upper-bound}
    For each edge $E$, suppose $a_i=f_i(\dots,\e_{i_k},\dots)$ and $b_i=f_i(\dots,\e_{i_k'},\dots)$ for $i\in[r]$. Let $l_i(\lambda)=a_i(1-\lambda)+b_i\lambda$, $i\in[r]$. Then
\begin{equation}\begin{aligned}
    T_E=\min\{&\max\{a_1,\dots,a_r\},\\
             &\max\{b_1,\dots,b_r\},\\
              &\text{for all }i\neq j,\begin{cases}
                \max_{k\in[r]}l_k((a_i-a_j)/(a_i+b_j-a_j-b_i)),&\text{if }a_i>a_j\text{ and }b_i<b_j,\\
                \max_{k\in[r]}l_k((a_i-a_j)/(a_i+b_j-a_j-b_i)),&\text{if }a_i<a_j\text{ and }b_i>b_j,\\
                1,&\text{otherwise}.
              \end{cases}\\
            \}&.
\end{aligned}\end{equation}
\end{lemma}

\begin{proof}
Essentially, $T_E$ is the maximum of $r$ linear functions $l_1,\dots,l_r$ in $\lambda$. Thus, its minimum on $[0,1]$ must be one of the following points:
\begin{itemize}
    \item the endpoint $\lambda=0$ with the minimum value $\max\{ l_1(0),\dots,l_r(0)\}$,
    \item the endpoint $\lambda=1$ with the minimum value $\max\{ l_1(1),\dots,l_r(1)\}$, or
    \item the points where $l_i$ intersects with $l_j$ for some $i\neq j$ at $\lambda^*$, with the minimum value $\max\{l_1(\lambda^*),\dots,l_r(\lambda^*)\}$.
\end{itemize}

For the last case, $l_i$ and $l_j$ intersect if and only if at the $l_i$, $l_j$ values have intersection on $\lambda\in[0,1]$, i.e., $a_i>a_j$ and $b_i<b_j$, or, $a_i<a_j$ and $b_i>b_j$. Then, under this condition, $l_i(\lambda^*)=l_j(\lambda^*)$ implies that $\lambda^*=(a_i-a_j)/(a_i+b_j-a_j-b_i)$. Note that $a_i=f_i(\dots,\e_{i_k},\dots)$ and $b_i=f_i(\dots,\e_{i_k'},\dots)$. Since all expressions are given by $a_i$ and $b_i$, $T_E$ can be expressed by $f_i$ values on the vertices of $E$ as in the lemma.
\end{proof}

Finally, we prove \Cref{thm:optimal-mixing-upper-bound}.

\begin{proof}[Proof of \Cref{thm:optimal-mixing-upper-bound}]
    By \Cref{lemma:T-E-upper-bound}, $T_E$ can be expressed by $f_i$ values on the vertices of $E$. Then, since there are only a fixed number of vertices (and hence edges) in $\mathcal{A}$, by \Cref{lemma:optimal-mixing-upper-bound}, we have that the upper bound
    \begin{equation}\min_{E\text{ is an edge of }\mathcal{A}}T_E=L^*\end{equation}
    can be expressed by $f_i$ values on the vertices of $\mathcal{A}$, which completes the proof.
\end{proof}

\begin{remark}\label{remark:optimal-mixing-two-player}
    For two-player games, there is an alternative logic encoding for the optimal mixing operation, which is implicitly used in \cite{daskalakisProgressApproximateNash2007} and \cite{kontogiannisPolynomialAlgorithmsApproximating2006}. The main idea as follows.
    \begin{itemize}
        \item First, we restrict the domain not to the edges of $\mathcal{M}$, but to the \emph{diagonal} of $\mathcal{M}$. That is, for strategy profile $(\x^1,\x^2)$ and $(\y^1,\y^2)$, we restrict the domain to $\{(\lambda\x^1+(1-\lambda)\y^1,\lambda\x^2+(1-\lambda)\y^2):\lambda\in[0,1]\}$.
        \item Then, instead of using linear upper bounds for the whole $f_i$ on the diagonal, we only relax the maximum term $\max_{(\x^1)'} u_1((\x^1)', \x^2)$ or $\max_{(\x^2)'} u_2(\x^1,(\x^2)')$ to a linear function. By doing so, $f_i$ is relaxed to a quadratic function $q_i$, and $f$ is relaxed to $\max\{q_1,q_2\}$.
        \item The minimum of $\max\{q_1,q_2\}$ on the diagonal can be explicitly expressed by $f_i$ values on the vertices of $\mathcal{M}$, which gives a different upper bound of the optimal mixing operation.
    \end{itemize}
    However, this encoding requires much more time to compute the final approximation guarantee than what we have shown in the main body. Thus, we do not use this encoding in the current LegoNE.
\end{remark}

\clearpage
\section{Extensions of LegoNE}\label{sec:extensions}

In this part, we show that the LegoNE framework can be extended to handle more general algorithmic problems.

\subsection{Example 1: ANE Algorithms in Polymatrix Games}\label{subsec:ANE-Polymatrix}

In the first example, we extend the LegoNE framework from $r$-player games to polymatrix games. 

Polymatrix games are a special kind of finite normal-form and graphical games. In a polymatrix game, the strategic interactions between players are captured by a graph $G = (V, E)$, where the nodes $V$ correspond to players, and edges $E$ indicate direct dependencies between players' utilities. Then, for any strategy profile $\x=(\x^{i})_{i\in V}$ of all players, the payoff function $u_i$ of player $i$ is only determined by the strategies of its neighbors $N(i)$ in the graph:
\begin{equation} u_i(\x)=(\x^{i})^\T\sum_{j\in N(i)} A_{ij}\ \x^j, \end{equation}
where $\x^j$ is the strategy of player $j$, and $A_{ij}$ is the payoff matrix between players $i$ and $j$. Besides, for the simplicity, in this part, for any two strategy profiles $\x$ and $\tilde{\x}$, we use the notation $(\x; \tilde{\x})$ to denote the strategy profile $(\x^i, \tilde{\x}^{-i})$ for all players $i$.

Since polymatrix games are also finite normal-form games, we can directly use the same loss-function representation as in $r$-player games. That is, for any strategy profile $\x$, we define the loss function for each player $i$ as
\begin{equation}
f_i(\x)=\max_{\tilde{\x}}u_i(\tilde{\x}; \x) - u_i(\x),
\end{equation}
and the approximation function $f(\x)=\max_i f_i(\x)$. 

The only known ANE algorithm for polymatrix games~\cite{deligkasComputingApproximateNash2017} has an approximation guarantee of $(1/2+\delta)$. We now show how to extend the LegoNE framework to prove this result. Following the LegoNE framework, we first give the Floyd-Hoare semantics of the ANE algorithm. Then, we use the instantiation and forgetting tactics to decide its validity in first-order logic over the reals (FOL-R), denoted $\mathsf{FOL}_\R$. 

For the first step, we need to encode an ANE algorithm into a logic formula. We consider the algorithm given by~\cite{deligkasComputingApproximateNash2017}. This algorithm uses a similar gradient-descent approach proposed in~\cite{tsaknakisOptimizationApproachApproximate2007} to compute and output the stationary point $\x^*$ of $f$. The encoding of the stationary point comes from equation (11) and definition (5) in \cite{deligkasComputingApproximateNash2017}:
\begin{equation}
    \begin{aligned}
        \forall\x'[&f(\x^*)\leq \max_{i} D f_i^\delta(\x^*;\x')+\delta\\
        &\wedge~ \forall i (Df_i^\delta(\x^*;\x')\leq \max_{\y} u_i(\y;\x')-u_i(\x^*;\x')-u_i(\x';\x^*)+u_i(\x^*))].
    \end{aligned}\label{eq:stationary-point-polymatrix-encoding}
\end{equation}
Here, $\delta$ is some fixed real constant.


Note that now the universal quantifier $\forall i$ is ranging over all players, instead of a fixed number $r$ of players.

For the second step, we show how to do the instantiation and forgetting. The key idea is, although there are unboundedly many players, we only use a finite number of approaches to handle them. The forgetting actually already occurs when we denote $(\x^i, \tilde{\x}^{-i})$ by $(\x;\tilde{\x})$, where we use a unified notation for all players $i$. We further do the forgetting to eliminate other occurrences of $i$ in the encoding:

\begin{itemize}
    \item  Denote functions $\max_i Df_i^\delta$ by $v$ and $u_i$ by $u$. By doing so, in notations $Df_i^\delta$ and $u_i$, the \emph{distinctions} between different players vanish.
    \item  After eliminating the occurrences of $i$, the derived upper bound of $Df_i^\delta(\x^*;\x')$ will hold for any $i$. Thus, it also holds for $\max_{i} Df_i^\delta(\x^*;\x')$, and we can replace it by $v(\x^*;\x')$.
\end{itemize}

Thus, after eliminating the occurrences of $i$ by forgetting, the encoding can be rewritten in the following form:
\begin{equation}\begin{aligned}
    \forall\x'[&f(\x^*)\leq v(\x^*;\x')+\delta\\ 
    &\wedge ~ v(\x^*;\x')\leq \max_{\y} u(\y;\x')-u(\x^*;\x')-u(\x';\x^*)+u(\x^*)].
\end{aligned}\end{equation}



Next, we instantiate $\x'$. The choice of $\x'$ is similar to that in LegoNE: $\x^*$, the stationary point, and $\overline{\x}$, the best response of all players against $\x^*$ (i.e., $\overline{\x}^i= \arg \max_{\x^i} u_i(\x^i,(\x^*)^{-i})$). Moreover, by the definition of best response, we have $\max_{\y} u(\y;\x^*)=u(\overline{\x};\x^*)$.

Thus, the instantiation of $\x'$ leads to the following formula:
\begin{equation}\begin{aligned}
    &f(\x^*)\leq v(\x^*;\overline{\x})+\delta \\
    \wedge~& v(\x^*;\x^*)\leq \max_{\y} u(\y;\x^*)-u(\x^*)-u(\x^*)+u(\x^*) \\
    \wedge~& v(\x^*;\overline{\x})\leq \max_{\y} u(\y;\overline{\x})-u(\x^*;\overline{\x})-u(\overline{\x};\x^*)+u(\x^*) \\
    \wedge~& \max_{\y} u(\y;\x^*)=u(\overline{\x};\x^*).
\end{aligned}\end{equation}
Here we use $u(\x^*;\x^*)=u(\x^*)$ by definition.

To be clearer, we can write it by substituting same terms by the same real variables to get a first-order real arithmetic formula:
\begin{equation}\begin{aligned}
    & a\leq b+\delta \wedge a\leq c+\delta\\
    \wedge~& b\leq d-e \wedge c\leq d-f-g+e\\
    \wedge~& d=g.
\end{aligned}\end{equation}

Using this formula and the inherent formulas (i.e., all variables are in $[0,1]$), we can again use cylindrical algebraic decomposition (CAD) to prove that the approximation guarantee of this algorithm is $(1/2+\delta)$, which can also be checked by humans. 

The above use of instantiation and forgetting is not dependent on any particular property of polymatrix games. Thus, it can actually be used for other kinds of graphical games (provided we have a proper programming language like LegoNE). Besides, the derivation above is easier than~\cite{deligkasComputingApproximateNash2017}, which instantiates the formula with $(\x+\overline{\x})/{2}$ instead, without justifying the choice of the coefficient $1/2$. However, under our framework, the choice of $1/2$ is also justified by the proof.

Finally, from a higher-level view, this example shows that by recursively applying forgetting and instantiation tactics, we can handle unboundedly many encodings that have a finite representation using indices.

\subsection{Example 2: Linear Programming Relaxations with Rounding Techniques}\label{subsec:LP-rounding}

In the second example, we show that LegoNE-style framework can be extended beyond ANE algorithms. Specifically, we consider the approximation analysis of algorithms designed using \emph{linear programming (LP) relaxations with rounding techniques}.

LP relaxations with rounding techniques are widely used to design polynomial-time approximation algorithms for combinatorial optimization problems~\cite{vaziraniApproximationAlgorithms2003}. The key idea is to first represent the problem as an integer linear program (ILP), then relax it to a linear program (LP), and finally round the solution of the LP to an integer solution~\cite{vaziraniApproximationAlgorithms2003}. As a concrete example, we consider the \emph{vertex cover} problem to illustrate how machines can prove the approximation ratios for its algorithm.

Given a graph $G=(V,E)$, a \emph{vertex cover} is a subset $V'\subseteq V$ such that each edge in $E$ has at least one endpoint in $V'$. The optimization problem is to find a minimum-size vertex cover. However, the decision version of vertex cover is $\mathsf{NP}$-complete~\cite{cormenIntroductionAlgorithms2009}. Thus, we have to resort to polynomial-time approximation algorithms.

To state such an algorithm, we first formulate the vertex cover problem as an integer linear program (ILP), denoted as $\mathsf{ILP}(V,E)$, formulated as follows:
\begin{equation}
\begin{aligned}
    \text{minimize} &\quad\sum_{v\in V} x_v\\
    \text{subject to} &\quad x_u+x_v\geq 1, \forall (u,v)\in E,\\
    &\quad  x_v\in\{0,1\}, \forall v\in V.
\end{aligned}\label{eq:vertex-cover-ILP}
\end{equation}
The optimal solution is a binary vector $x^*\in\{0,1\}^{V}$, corresponding to a vertex cover, where $x^*_v=1$ if and only if $v$ is in the cover. 

Then we present the desired algorithm $\Gamma$ with two steps as follows:
\begin{enumerate}
    \item $x= \sol(\mathsf{LP}(V,E))$: 
    We relax the ILP to a linear program (LP) by relaxing the constraints $x_v\in\{0,1\}$ to $0\leq x_v\leq 1$. The LP is denoted as $\mathsf{LP}(V,E)$. Solve $\mathsf{LP}(V,E)$ and we can get a fractional solution $x$.

    \item $x'=\mathsf{Round}(x)$: We 
    then round $x$ to an integer solution $x'$ by setting $x'_v=1$ if $x_v\geq 1/2$, and $=0$ otherwise. Output $x'$ as the solution.
\end{enumerate}

It was shown that the output $x'$ of this algorithm is indeed a vertex cover and the approximation ratio of algorithm $\Gamma$ is $2$~\cite{vaziraniApproximationAlgorithms2003}, i.e., the size of the output is at most twice the size of the optimal solution. Now, let us see how the LegoNE framework can be extended for the approximation analysis of this algorithm. Here, we do not consider how to prove the correctness (i.e., Algorithm $\Gamma$ is indeed a polynomial-time vertex-cover algorithm) by machines. Again, we need to
\begin{enumerate*}
    \item give the Floyd-Hoare semantics of the algorithm;
    \item use the instantiation and forgetting tactics to decide its validity in $\mathsf{FOL}_\R$.
\end{enumerate*}

For the first step, just like ANE algorithms, algorithm $\Gamma$ can be seen as a combination of two building blocks: $\mathsf{LP}$ and $\mathsf{Round}$, and we can encode them as follows:
\begin{itemize}
    \item Since we relax the constraints, the optimal objective value $\sum_{v \in V} x_v $ of $\mathsf{LP}(V,E)$ is no more than the optimal objective value of $\mathsf{ILP}(V,E)$. 
Denote the optimal solution to $\mathsf{ILP}(V,E)$ as $x^*$, the encoding of $x=\sol(\mathsf{LP}(V,E))$ is:
    \begin{equation} \left(\sum_{v \in V} x_v \leq \sum_{v \in V} x_v^*\right).\end{equation}
   
    
    \item By simple reasoning, the encoding of $x'=\mathsf{Round}(x)$ is:    
    \begin{equation}\forall v (x_v' \leq 2 x_v).\end{equation}

    \item We also have an inherent encoding of $x^*$ as the optimal solution of $\mathsf{ILP}(V,E)$:
    \begin{equation}\forall y\left(\sum_{v\in V} x_v^*\leq \sum_{v\in V} y_v\right),\end{equation}
    where the domain of $y$ is all feasible solutions of $\mathsf{ILP}(V,E)$.
\end{itemize}

For the second step, we encounter a problem that resembles those in \Cref{subsec:ANE-Polymatrix}: We have unboundedly many variables and constraints, but they are finitely represented using indices. Our solution is similar. Facing notation $x_v$, we forget the information given by subscript $v$ and treat $x_v$ equally for all $v$'s. 

Bearing this idea, we can present the instantiation tactics on $\forall y$ and $\forall v$:
\begin{itemize}
    \item For $y$, its value can only be either $x'$ or $x^*$. Other choices are unlikely to produce.
    \item For $v$, to eliminate the quantifier $\forall v$, we have a different approach from \Cref{subsec:ANE-Polymatrix}. Here, by the characteristic of the sum operator, we can handle it by summing up over all $v$'s, i.e., from premise $\forall v(f(v)\leq g(v))$ to derive a conclusion $\sum_v f(v)\leq \sum_v g(v)$. In this way, the quantifier $\forall v$ is eliminated.
\end{itemize}

Following the above elimination deductions, we can derive 
\begin{equation} \sum_{v\in V} x_v'\leq \sum_{v\in V} (2x_v)=2\sum_{v\in V} x_v\end{equation}
from $\forall v (x_v' \leq 2 x_v)$, and
\begin{equation}\sum_{v\in V} x_v^*\leq \sum_{v\in V} x'_v\wedge \sum_{v\in V} x^*_v\leq \sum_{v\in V} x^*_v\end{equation}
from $\forall y(\sum_v x_v^*\leq \sum_v y_v)$.
We also know from the encoding of $x=\sol(\mathsf{LP}(V,E))$ that
\begin{equation}\sum_{v\in V} x_v\leq \sum_{v\in V} x_v^*.\end{equation}
The goal of the proof is to show that
\begin{equation}\sum_{v\in V} x'_v\leq 2\sum_{v\in V} x^*_v.\end{equation}

Again, we can use the forgetting tactic. We can treat sum terms as whole notations. By doing so, all derived inequalities and the proof goal only involves real-valued terms, and need to show:
\begin{equation} (a\leq 2b\wedge c\leq a\wedge c\leq c\wedge b\leq c)\to a\leq 2c.\end{equation}
This problem also reduces to deciding the validity of a formula in $\mathsf{FOL}_\R$. We can again use CAD to solve this problem. 

Importantly, the above instantiation and forgetting procedure does not depend on the specific structure of the vertex cover problem. It only concerns the form of ILP and our treatment on indices. Thus, the above procedure can be applied to other linear programming relaxations with rounding techniques.

\end{document}